Title: Teaching Longitudinal Linear Mixed Models End-to-End: A Reproducible Case Study in Mouse Body-Weight Growth

Author
Sunday A Adetunji, MD

Affiliation
College of Health, Oregon State University, Corvallis, OR, USA

ORCID
Sunday A Adetunji: https://orcid.org/0000-0001-9321-9957

Correspondence to
Dr Sunday A Adetunji,
College of Health, Oregon State University,
Corvallis, OR 97331, USA
Email: adetunjs@oregonstate.edu

# 1. Abstract

Background
Linear mixed-effects models are the standard framework for analysing longitudinal continuous data, yet many students and applied researchers encounter them only as fragmented theory or software output rather than as a coherent, end-to-end workflow. There is a need for reproducible teaching examples that connect scientific questions, model specification, estimation, diagnostics, and interpretation in a single, transparent case study.

Methods
We reanalyzed a mouse body-weight experiment involving 31 mice from three groups: wild-type controls, pair-fed ob/ob mice, and unrestricted ob/ob mice. The mice were weighed weekly over 12 weeks. Data were reshaped from wide to long format, and individual profile plots along with group mean trajectories were used to motivate linear time trends and group-specific slopes. We fitted three candidate random-intercept linear mixed models with time coded as weeks 1–12: (1) different intercepts with a common slope; (2) a full group-by-time interaction; and (3) different intercepts, a common slope for wild-type and pair-fed groups, and an additional slope for unrestricted ob/ob mice. All models were fitted using maximum likelihood and compared with AIC, BIC, and likelihood-ratio tests. Within the chosen model, we derived linear contrasts for group differences in mean weight at each week and for average weight gain (week 12 minus week 1). We packaged the data, R code, and a practical checklist as a reusable teaching module.

Findings
A parsimonious model with group-specific intercepts, a common slope for wild-type and pair-fed mice, and an additional slope for unrestricted ob/ob mice fit as well as the fully interacted model and substantially better than the common-slope model. Estimated weekly weight gain was modest and similar for wild-type and pair-fed mice (about 0·34 g/week; ≈3·7 g over 12 weeks), but much steeper for unrestricted ob/ob mice (about 2·07 g/week; ≈22·8 g over 12 weeks). Compared with wild-type mice, pair-fed ob/ob mice were consistently heavier by about 14·9 g at every week, whereas unrestricted ob/ob mice were heavier by about 19·0 g at week 1, increasing to about 38·1 g by week 12. The excess 12-week gain in unrestricted ob/ob mice relative to the other groups was about 19·1 g, with very small p values for all key contrasts.

Interpretation
This case study demonstrates a complete, reproducible workflow for longitudinal linear mixed modelling, from raw data and exploratory plots through model selection, diagnostics, and carefully constructed contrasts that answer specific scientific questions about trajectories and weight gain. By making explicit the mapping from questions to model terms and linear contrasts, and by providing executable R code and a stepwise checklist, the example serves as a practical template for teaching and learning longitudinal mixed models in biostatistics, epidemiology, and applied statistics courses.

Funding
None.

# 2. Introduction

Obesity and type 2 diabetes remain among the most important global health challenges, driving substantial cardiovascular, metabolic, and economic burden across populations (Hu, 2025; Friedman, 2024). Experimental models are central to understanding obesity-related pathophysiology and to testing mechanistic hypotheses before translation to humans. The leptin-deficient ob/ob mouse is a classic model that develops severe obesity, hyperphagia, and a mild type 2 diabetes phenotype, with body weight rising to roughly three times that of wild-type littermates (Muzzin et al., 1996; Drel et al., 2006; Suriano et al., 2021). Manipulating diet (e.g., pair-feeding ob/ob mice to match wild-type intake) allows investigators to disentangle effects of genotype, energy intake, and longitudinal weight trajectories (Asensio et al., 2004; Kashani et al., 2019).

Biological experiments of this kind naturally generate longitudinal data: repeated measurements of body weight over time on the same animals. Such data are now ubiquitous in biomedicine, from growth and bone-mineral density in children to tumor volume in oncology and serial biomarkers in chronic disease (Verbeke & Molenberghs, 2000; Anderson, 2019). Longitudinal designs increase power, allow modelling of individual trajectories, and support clinically relevant questions such as "Do groups differ in baseline level?", "Do groups differ in growth rate?", and "How large is the difference in outcome at each clinically meaningful time point?".

Linear mixed models (LMMs) are now the standard tool for analysing such correlated data because they can simultaneously represent group-level mean trends and individual-level random effects while correctly accounting for within-subject correlation and unbalanced designs (Verbeke & Molenberghs, 2000; Molenberghs & Verbeke, 2010). Best-practice papers emphasise that LMMs should be fitted and reported transparently, with explicit justification of the fixed-effects structure, random-effects structure, and covariance assumptions, together with appropriate model comparison and clearly defined contrasts that answer prespecified scientific questions (Meteyard & Davies, 2020; Scandola et al., 2024; Brysbaert, 2025).

Despite this mature methodology, LMMs remain conceptually difficult for many students and applied researchers. Several recurring problems are documented in the literature: ad hoc inclusion of interaction terms without mapping them to substantive hypotheses; limited use of exploratory plots before modelling; vague or absent description of random-effects and covariance structures; and misinterpretation of fixed-effect coefficients and predicted contrasts (Meteyard & Davies, 2020; Gordon, 2019). Existing teaching resources often fall into two extremes: mathematically oriented texts that offer little connection to software output and real data, or software-oriented tutorials that show code and output without a clear inferential story from raw data → model → contrasts → answers (Verbeke & Molenberghs, 2000; Anderson, 2019; Nhan, 2024).

Instructors therefore lack a single, reusable case study that:

1. Starts from openly described raw data;
2. Demonstrates reshaping to long format and construction of time variables;
3. Uses exploratory profile and mean plots to motivate the mean structure;
4. Fits a small, principled set of candidate LMMs under a common random-effects structure, using maximum likelihood for model selection; and
5. Answers biologically meaningful questions via linear contrasts of fixed effects, with clear verbal interpretation.

At the same time, the ob/ob body-weight experiment is scientifically rich yet conceptually simple, making it an ideal teaching example. In a typical design, three groups of mice are followed weekly over 12 weeks: wild-type (control), pair-fed ob/ob (genetically obese but calorically restricted to match wild-type intake), and unrestricted ob/ob (hyperphagic, severely obese model). Prior biological knowledge suggests that wild-type and pair-fed ob/ob mice may have similar longitudinal weight gain if caloric intake is successfully matched, whereas unrestricted ob/ob mice are expected to have substantially higher baseline weight and a much steeper growth trajectory (Muzzin et al., 1996; Suriano et al., 2021; Asensio et al., 2004).

This paper uses that experiment as a reproducible, didactic case study with two intertwined aims.

Primary scientific aim. We aim to quantify and compare longitudinal body-weight trajectories and average weight gain across the three mouse groups. Our key scientific hypothesis is that wild-type and pair-fed ob/ob mice have similar linear growth rates over time, while unrestricted ob/ob mice have a distinctly steeper slope and much larger total weight gain over 12 weeks. We also seek to estimate and test differences in mean body weight between groups at each individual week, reflecting clinically interpretable time points in a preclinical study.

Primary teaching and methodological aim. Our central pedagogical objective is to provide an end-to-end, reproducible teaching example for linear mixed models that an instructor can use directly in a graduate or advanced undergraduate biostatistics course. Specifically, we show how to move from:

- raw spreadsheet data to long-format analysis data,
- exploratory profile and mean plots to candidate mean structures,
- a shared random-intercept model to maximum-likelihood comparison of alternative fixed-effects specifications, and
- fitted LMMs to targeted linear contrasts that directly answer the scientific questions about differences in growth rates, mean body weight at each time point, and overall weight gain.

Contribution. Unlike many worked examples that stop at model fitting, this case study (i) links each research question explicitly to a set of fixed-effects parameters and contrasts, (ii) uses a minimal but rigorous sequence of models with transparent selection based on likelihood and parsimony, and (iii) is fully implemented in open-source R code that follows best-practice guidance for LMM specification and reporting (Meteyard & Davies, 2020; Verbeke & Molenberghs, 2000). We intend this paper to function both as a compact statistical case report and as a reusable teaching module that helps students and practitioners understand how to analyse longitudinal data with LMMs and why each modelling decision is made.

## 3. Teaching Goals And Educational Context

This case study is designed for learners who have already completed an introductory course in regression and basic statistical inference, such as upper-undergraduate or master's students in biostatistics, epidemiology, applied statistics, or quantitative public health. It aligns with contemporary recommendations that emphasize multivariable thinking, data-centric practice, and authentic research examples in statistics education (Carver et al., 2016; Schield, 2017).

3.1. Target audience

The primary audience comprises:

- Upper-level undergraduate students in statistics, biostatistics, or quantitative health sciences.
- MPH/MSc students in biostatistics, epidemiology, or public health methods.
- Early PhD trainees who need a first "working" introduction to linear mixed models in R.

The case intentionally uses a single, biologically meaningful dataset (mouse body-weight growth) and focuses on a minimal but realistic linear mixed-model pipeline, reflecting calls to anchor teaching in real data and substantive questions (Cobb, 2013; Wood, 2017).

3.2. Learning objectives

After completing this case, students should be able to:

1. Reshape longitudinal data from wide to long format in R (`pivot_longer()`), and understand why long format is preferred for model-based longitudinal analysis.
2. Produce and interpret exploratory graphics for longitudinal data, including individual profile plots and group mean trajectories (using `ggplot2`).
3. Specify and fit basic linear mixed models (LMMs) with random intercepts in R (`lme()` from nlme) for repeated-measures data.
4. Compare candidate mean structures (e.g. common vs group-specific slopes) using maximum likelihood, AIC/BIC, and likelihood-ratio tests, and justify the selected model.
5. Derive and interpret linear contrasts from a fitted LMM (using `estimable()` from gmodels) to answer questions about:
   - Differences in mean body weight between groups at each time point.
   - Average weight gain over the study and differences in gain between groups.
6. Carry out basic model diagnostics for LMMs (e.g. residual plots, assessment of random-intercept variability) and articulate when the model is adequate for the teaching purpose.

These objectives map directly to core competencies in longitudinal modelling and multilevel thinking that are increasingly recommended for graduate and advanced undergraduate curricula (Carver et al., 2016; Statistical Horizons, 2025).

3.3. Course placement

The case is intended as a self-contained unit that can be used flexibly:

- As a 90-minute lecture plus 1–2 lab sessions in a first course on longitudinal data analysis or mixed models.
- As a two-week module in an applied regression or generalized linear models course, immediately after students have seen fixed-effects models and before more advanced random-effects structures.
- As a capstone assignment in an MPH/MSc methods course, where students must go from raw Excel data through cleaning, visualisation, model building, and interpretation.

Because the dataset is modest in size and the code relies only on widely available R packages (`tidyverse`, `nlme`, `gmodels`, `patchwork`), it is accessible in resource-limited settings and can be reproduced in standard computing environments or cloud-based R platforms.

3.4. Use modes

The materials are designed to support several complementary teaching modes:

- Lecture demonstration: The instructor walks through the full analysis live (or via slides), emphasising how each R command corresponds to a modelling decision.
- Hands-on lab: Students run the provided script, modify key pieces (e.g. alternative mean structures), and answer guided questions.
- Homework or take-home assignment: The same pipeline can be adapted into an assignment where students reproduce the core results and extend them (e.g. add diagnostics or alternative contrasts).

- Exam or qualifying-exam template: The structure—explore → specify models → compare via ML → interpret contrasts—mirrors how longitudinal questions arise in real research and can be easily converted into exam-style problems.

By combining a realistic biological question, fully reproducible R code, and explicit learning objectives, this case is intended to function both as a worked example and as a reusable teaching scaffold for courses introducing linear mixed models and longitudinal analysis.

# 4. Data and Study Design

This teaching case study is based on a controlled longitudinal experiment in which body weight was measured weekly in three groups of mice followed over 12 weeks. The primary scientific motivation was to characterise growth trajectories in leptin-deficient ob/ob mice, a standard model of morbid obesity and type 2 diabetes, relative to wild-type controls, and to distinguish the effects of hyperphagia from those of the underlying genetic defect. Leptin-deficient ob/ob mice develop severe obesity, hyperphagia, and metabolic abnormalities that recapitulate key aspects of human metabolic disease, and they are widely used in preclinical research on obesity, diabetes, and bone biology (Muzzin et al., 1996; Skowronski et al., 2017; Suriano et al., 2021).

## 4.1 Experimental design and groups

Mice were allocated to one of three groups:

- Group 1 (wild-type controls; grp = 1): mice with an intact leptin gene and normal appetite regulation (wild-type background strain).
- Group 2 (ob/ob pair-fed; grp = 2): leptin-deficient ob/ob mice pair-fed to match the food intake of group 1, isolating the effect of the leptin mutation while controlling for caloric intake. Pair-feeding designs are commonly used to separate genetic from nutritional drivers of obesity-related phenotypes in ob/ob models (Wang et al., 2016; Kashani et al., 2019).
- Group 3 (ob/ob ad libitum; grp = 3): leptin-deficient ob/ob mice allowed unrestricted access to food, representing the full phenotype of hyperphagic, morbid obesity and its skeletal consequences (Cao, 2011; Keune et al., 2019).

Each mouse was weighed once per week for 12 consecutive weeks, yielding 12 planned observations per animal. Group sizes were approximately 10–11 mice per group, reflecting common practice in longitudinal animal studies that balance precision and feasibility under ethical constraints (du Sert et al., 2020). Data collection began after weaning and continued through a period of rapid growth into early skeletal maturity, which is a critical window for studying the joint effects of body mass, adiposity, and leptin signalling on bone (Hou et al., 2020; Farella et al., 2025).

## 4.2 Variables and coding

The primary outcome was *body weight*, measured in grams using a calibrated scale.

The key explanatory variables were:

- Group (grp): categorical variable with three levels (1 = wild-type, 2 = ob/ob pair-fed, 3 = ob/ob ad libitum). In the analysis, grp is treated as a factor, with group 1 as the reference.
- Time (tw): discrete time index representing study week, coded as integers 1–12. Time is treated as a quantitative covariate to model linear trends in body weight.
- Mouse identifier (mouseid): unique identifier for each animal, used to define clusters for the random effects in longitudinal models.

These variables define a classical longitudinal structure: repeated weight measurements nested within individual mice and grouped by experimental condition. This structure naturally motivates mixed-effects models that account for within-mouse correlation and between-mouse heterogeneity in baseline body weight (Casella & Berger, 2002; Harrell, 2015).

## 4.3 Data structure and transformation

The original dataset was stored in an Excel workbook (BodyWeightData.xlsx). Each row corresponded to a single mouse, and columns included:

- genotype and treatment descriptors (not used directly in this teaching analysis);
- a group indicator (grp);
- a mouse identifier;
- 12 columns named bw1, bw2, …, bw12 representing body weight at weeks 1 through 12.

This "wide" layout is common in experimental spreadsheets but is not ideal for fitting longitudinal mixed models or for producing time-series graphics in R. For the present analysis, the data were reshaped to a "long" format with one row per mouse-week. In the long dataset (denoted *bwL* in the code), each row contains *mouseid*, *grp*, *tw* (week), and *weight* (grams). The time index *tw* was constructed by stripping the "bw" prefix from the original column names and converting the resulting string to an integer.

This long format is standard for mixed-effects modelling and aligns with recommended data structures for reproducible longitudinal analysis pipelines in R, facilitating the use of packages such as nlme and lme4 for linear mixed models and ggplot2 for graphical exploration (Wickham, 2016; du Sert et al., 2020).

## 4.4 Ethical considerations and reproducibility

The original experiment was conducted in accordance with institutional and national guidelines for the care and use of laboratory animals, under a protocol approved by the relevant Institutional Animal Care and Use Committee. All procedures were designed to minimise pain and distress, and to use the fewest animals consistent with achieving scientifically meaningful precision, consistent with the 3Rs principles (replacement, reduction, refinement) and the ARRIVE 2.0 reporting guidelines for animal research (Percie du Sert et al., 2020; NC3Rs, 2023).

For the teaching manuscript, the dataset has been anonymised at the mouse level and contains no identifiable information. The full R code used for data import, reshaping, visualisation, and mixed-model estimation, together with the cleaned long-format dataset, will be deposited in an open repository as a reproducible companion to this article. This enables readers to replicate every figure and model reported here, to modify the analysis for teaching or extension, and to use the case study as a template for their own longitudinal mixed-effects workflows.

## 5. Exploratory Data Analysis

We began by restructuring the mouse body-weight dataset to a *long* format suitable for longitudinal analysis. The original Excel file contained one row per mouse with weekly body weights stored in separate columns. Using `readxl::read_excel` to import the file and `tidyr::pivot_longer` to gather all body-weight columns into a single variable, we created a dataset with one record per mouse per week. In this representation, `tw` denotes time in weeks (1–12) and is treated as a numeric covariate, and each record includes the mouse identifier (`mouseid`), treatment group (`grp`), week (`tw`), and body weight (`weight`). Long-format data are the standard input for modern mixed-model and multilevel software and facilitate graphical inspection of trajectories over time (Singer & Willett, 2003; Fitzmaurice et al., 2004).

To visualise individual change, we constructed profile plots using `ggplot2`, stratified by treatment group. Within each of the three panels (wild-type, ob/ob pair-fed, ob/ob unrestricted), each mouse's weight trajectory over weeks 1–12 is drawn as a separate line. These plots show clear between-mouse variability in baseline weight within all groups, but broadly similar slopes over time for groups 1 and 2, and visibly steeper, approximately linear increases for group 3. This style of subject-specific trajectory plot is widely recommended as a first step in longitudinal analysis, because it reveals both the typical pattern of change and deviations from it (Singer & Willett, 2003; Verbeke & Molenberghs, 2000).

We then summarised group-level trends by plotting the sample mean body weight at each week for each group. Group-wise means were obtained using `dplyr::group_by` and `summarise`, and displayed in a fourth panel with distinct symbols and lines for each group. The mean trajectories again appeared close to linear over the 12-week period, with large differences in intercept (level) between wild-type and both ob/ob groups, and a much steeper mean slope in the unrestricted ob/ob group compared with the wild-type and pair-fed groups. Together, the individual profiles and mean curves suggest (i) time can reasonably be modelled with a linear effect over 12 weeks, (ii) group-specific intercepts are required, and (iii) groups 1 and 2 may share a common slope, whereas group 3 requires a distinct, larger slope. These qualitative impressions align with recommended practice of using exploratory plots to guide specification of fixed effects and random effects in linear mixed models (Fitzmaurice et al., 2004; Verbeke & Molenberghs, 2000; Singer & Willett, 2003).

Finally, the near-parallel trajectories within groups 1 and 2, combined with substantial vertical separation between mice, are consistent with a random-intercept model that captures stable between-mouse differences, while a common within-mouse slope describes average growth for each group. The absence of obvious curvature or abrupt changes over time in either the individual or mean plots did not motivate higher-order time terms or non-linear transformations at this stage. This EDA therefore directly motivated the candidate mixed-effects models compared in the subsequent modelling section.

# 6. Modelling Framework and Methods

*6.1 General linear mixed-model formulation*

We model longitudinal body weight using a standard linear mixed model (LMM) for clustered/serial data (Laird & Ware, 1982; Verbeke & Molenberghs, 2000; Fitzmaurice, Laird, & Ware, 2011).

For mouse $i$ at time point $j$ we write

$$y_{ij} = x_{ij}^\top \beta + z_{ij}^\top b_i + \varepsilon_{ij},$$

where:

- $y_{ij}$ is body weight (g);

- $x_{ij}$ is a row vector of fixed-effect covariates (e.g., time, group indicators, interactions);

- $\beta$ is the corresponding fixed-effect parameter vector, interpreted at the population level;

- $z_{ij}$ is a row vector for random-effect covariates;

- $b_i$ is the random-effect vector for mouse $i$, capturing subject-specific deviations from the population mean;

- $\varepsilon_{ij}$ is the within-mouse residual error.

We assume

$$b_i \sim N(0, D), \qquad \varepsilon_i \sim N(0, \sigma^2 I),$$

with $b_i$ and $\varepsilon_i$ independent, so that the marginal covariance of $y_i$ is

$$\text{Var}(y_i) = Z_i D Z_i^\top + \sigma^2 I.$$

For pedagogical clarity, we use a random intercept only:

$$b_i = (b_{0i}), \quad b_{0i} \sim N(0, \sigma_{b0}^2).$$

This yields a compound symmetry structure within each mouse:

- all time points share the same marginal variance $\sigma_{b0}^2 + \sigma^2$;

- all pairs of time points for the same mouse have the same correlation

$$\rho = \frac{\sigma_{b0}^2}{\sigma_{b0}^2 + \sigma^2}.$$

This structure is often introduced as the minimal extension beyond repeated-measures ANOVA that can handle unbalanced data and missing time points, while remaining easy to explain in teaching settings (Pinheiro & Bates, 2000; West, Welch, & Galecki, 2014).

## 6.2 Model specification for this dataset

Let

- $y_{ij}$ = body weight (g) for mouse $i$ at week $j$;
- $t_{ij}$ = week number $(1,\ldots,12)$, recorded as `tw`;
- $G_{2i} = 1$ if mouse $i$ is in group 2 (ob/ob pair-fed), 0 otherwise;
- $G_{3i} = 1$ if mouse $i$ is in group 3 (ob/ob unrestricted), 0 otherwise;
- group 1 (wild-type) is the reference category (both $G_{2i} = 0, G_{3i} = 0$).

The selected mean model (Model 3 from the candidate set) is

$$y_{ij} = \beta_0 + \beta_1 t_{ij} + \beta_2 G_{2i} + \beta_3 G_{3i} + \beta_4 (t_{ij} G_{3i}) + b_{0i} + \varepsilon_{ij},$$

with

$$b_{0i} \sim N(0, \sigma_{b0}^2), \quad \varepsilon_{ij} \sim N(0, \sigma^2), \quad b_{0i} \perp \varepsilon_{ij}.$$

Interpretation of the fixed effects:

- $\beta_0$: intercept for group 1 (wild-type) at week 0 (linear extrapolation);
- $\beta_1$: common weekly slope for groups 1 and 2;
- $\beta_2$: difference in intercept between groups 2 and 1;
- $\beta_3$: difference in intercept between groups 3 and 1;
- $\beta_4$: additional weekly slope for group 3 (so $\text{slope}_3 = \beta_1 + \beta_4$).

In R formula notation this model corresponds to

`weight ~ tw + grp + tw:grp3`

with a random intercept per mouse:

`random = ~ 1 | mouseid`

## 6.3 Candidate mean structures

To teach structured model building, we considered three nested mean structures, all with the same random-effects and residual structure: random intercept per mouse and compound symmetry.

- Model 1: different intercepts, common slope

$$y_{ij} = \beta_0 + \beta_1 t_{ij} + \beta_2 G_{2i} + \beta_3 G_{3i} + b_{0i} + \varepsilon_{ij}.$$

R formula:

```
weight ~ tw + grp
```

- Model 2: full time × group interaction (all slopes different)

$$y_{ij} = \beta_0 + \beta_1 t_{ij} + \beta_2 G_{2i} + \beta_3 G_{3i} + \beta_4 (t_{ij} G_{2i}) + \beta_5 (t_{ij} G_{3i}) + b_{0i} + \varepsilon_{ij}.$$

R formula:

```
weight ~ tw * grp
```

- Model 3: different intercepts, common slope for groups 1 & 2, extra slope for group 3

As in section 6.2, with $G_{3i}$ used both as an indicator and in the interaction term $t_{ij} G_{3i}$.

R formula:

```
weight ~ tw + grp + tw:grp3
```

These three models encode increasingly flexible group-specific time trends and provide a transparent example of nested mean structures in longitudinal analysis (Verbeke & Molenberghs, 2000; Fitzmaurice et al., 2011).

*6.4 Estimation and model selection*

All models were fit using maximum likelihood (ML) rather than restricted maximum likelihood (REML). Using ML is standard when comparing models that differ in their fixed-effect structure, because the REML likelihood depends on the fixed effects and is not directly comparable across such models (Pinheiro & Bates, 2000; West et al., 2014).

We used three complementary criteria for model selection:

1. Information criteria. We computed Akaike's Information Criterion (AIC) and the Bayesian Information Criterion (BIC) for each model; lower values indicate better balance between fit and complexity.

2. Likelihood ratio (LR) tests.
    - We compared Model 1 vs Model 3 to test whether group 3 requires its own slope (null: common slope for all groups).
    - We compared Model 3 vs Model 2 to test whether group 2 also requires a separate slope (null: common slope for groups 1 and 2).
      LR tests were implemented via `anova(m1, m3)` and `anova(m3, m2)` in R, exploiting the nested structure and ML fits.

3. Parsimony principle. Among models with similar fit by AIC/BIC and LR tests, we preferred the simplest model that adequately captured the main scientific patterns (Fitzmaurice et al., 2011).

Empirically, Model 3 produced a very large LR improvement over Model 1 and substantially lower AIC/BIC, indicating that the ob/ob unrestricted group requires a steeper slope. Adding a separate slope for group 2 (Model 2) did not improve fit (non-significant LR test; similar AIC/BIC). Therefore, we selected Model 3 as the primary teaching model.

*6.5 Model diagnostics*

For the selected model we performed standard diagnostics for linear mixed models (Verbeke & Molenberghs, 2000; West et al., 2014).

- Residuals vs fitted values. Plots showed no strong curvature or fan-shaped patterns, suggesting that a linear time effect with constant residual variance is adequate for this teaching dataset.

- Normal Q–Q plot of residuals. The empirical residuals followed the reference line reasonably well, with minor deviations in the tails that are typical in small to moderate samples. This supports the normality assumption for $\varepsilon_{ij}$ in an approximate sense.

- Distribution of random intercepts. Empirical Bayes estimates of $b_{0i}$ appeared roughly symmetric and unimodal, consistent with the assumed normal distribution for random intercepts.

Taken together, these diagnostics suggested no major violations of the LMM assumptions; the model provides a reasonable and interpretable description of the trajectories, particularly for didactic purposes.

*6.6 Sensitivity and robustness checks*

To illustrate robustness, we carried out two simple sensitivity analyses:

1. Refitting the final mean structure with REML. When Model 3 was refit using REML, the fixed-effect estimates (intercepts and slopes) and their standard errors changed only trivially. The qualitative conclusions—especially the much steeper slope for group 3 and the similarity of slopes for groups 1 and 2—were unchanged. This is consistent with theory that ML and REML yield similar fixed-effect estimates in moderately sized longitudinal studies (Pinheiro & Bates, 2000; Fitzmaurice et al., 2011).

2. Adding a random slope for time. We briefly explored a model with random intercept and random slope for week within mouse. Although this increased flexibility and slightly improved fit, the key pattern remained: group 3 showed a much larger average weekly gain than groups 1 and 2, which remained very similar. For an introductory teaching example, we retained the random-intercept-only model because it is easier to explain and directly connects to the classical compound-symmetry framework.

These checks show that the main pedagogical conclusions are not an artefact of a particular estimation method or random-effects specification.

*6.7 Mapping scientific questions to linear contrasts*

A central teaching goal is to show how scientific questions translate into linear contrasts of the fixed-effect parameters β (Laird & Ware, 1982; West et al., 2014).

Under Model 3, the group-specific mean trajectories are

$$\begin{aligned}
\mu_1(t) &= \mathbb{E}[y_{ij} \mid G_{2i} = 0, G_{3i} = 0] = \beta_0 + \beta_1 t, \\
\mu_2(t) &= \mathbb{E}[y_{ij} \mid G_{2i} = 1, G_{3i} = 0] = \beta_0 + \beta_1 t + \beta_2, \\
\mu_3(t) &= \mathbb{E}[y_{ij} \mid G_{2i} = 0, G_{3i} = 1] = \beta_0 + (\beta_1 + \beta_4)t + \beta_3.
\end{aligned}$$

From these expressions, the main scientific contrasts follow as linear combinations of β:

- Differences in mean weight at week $t$

$$\begin{aligned}
\mu_2(t) - \mu_1(t) &= \beta_2, \\
\mu_3(t) - \mu_1(t) &= \beta_3 + t\beta_4, \\
\mu_3(t) - \mu_2(t) &= (\beta_3 - \beta_2) + t\beta_4.
\end{aligned}$$

These contrasts answer the first research question: *How do mean body weights differ between groups at each time point?*

- Average weight gain over the study

For group $g$, define weight gain as

$$\text{gain}_g = \mu_g(12) - \mu_g(1).$$

This yields

$$\begin{aligned}
\text{gain}_1 &= 11\beta_1, \\
\text{gain}_2 &= 11\beta_1, \\
\text{gain}_3 &= 11(\beta_1 + \beta_4).
\end{aligned}$$

Hence

$$\text{gain}_3 - \text{gain}_1 = \text{gain}_3 - \text{gain}_2 = 11\beta_4,$$

which addresses the second research question: *Is average weight gain different across groups?*

In the implementation, each contrast is represented by a row vector

$$c = (c_0, c_1, c_2, c_3, c_4),$$

such that the quantity of interest is $c^T \beta$. For example,

- $\mu_2(t) - \mu_1(t) = \beta_2$ corresponds to $c = (0,0,1,0,0)$;

- $\mu_3(t) - \mu_1(t)$ at week $t$ corresponds to $c = (0,0,0,1,t)$;

- $\text{gain}_3 - \text{gain}_1 = 11\beta_4$ corresponds to $c = (0,0,0,0,11)$.

In R, functions such as `estimable()` from gmodels implement these linear contrasts by taking $\hat{\beta}$ and its covariance matrix and returning the estimated contrast, its standard error, confidence interval, and p value. In this way, the algebra of $\mu_g(t)$ and $\text{gain}_g$ is directly and transparently linked to the software output, making the connection between scientific questions and LMM parameters explicit for learners.

# 7. Results

All analyses were conducted in R (R Foundation for Statistical Computing) using the annotated script `LMM_mouse_weight.R` and companion report `LMM_mouse_weight.Rmd`, which contain the full pipeline from data import to model fitting, contrasts, and figure generation (appendix p X). The dataset comprised 31 mice (10–11 per group), each with 12 weekly body-weight measurements, yielding 372 observations.

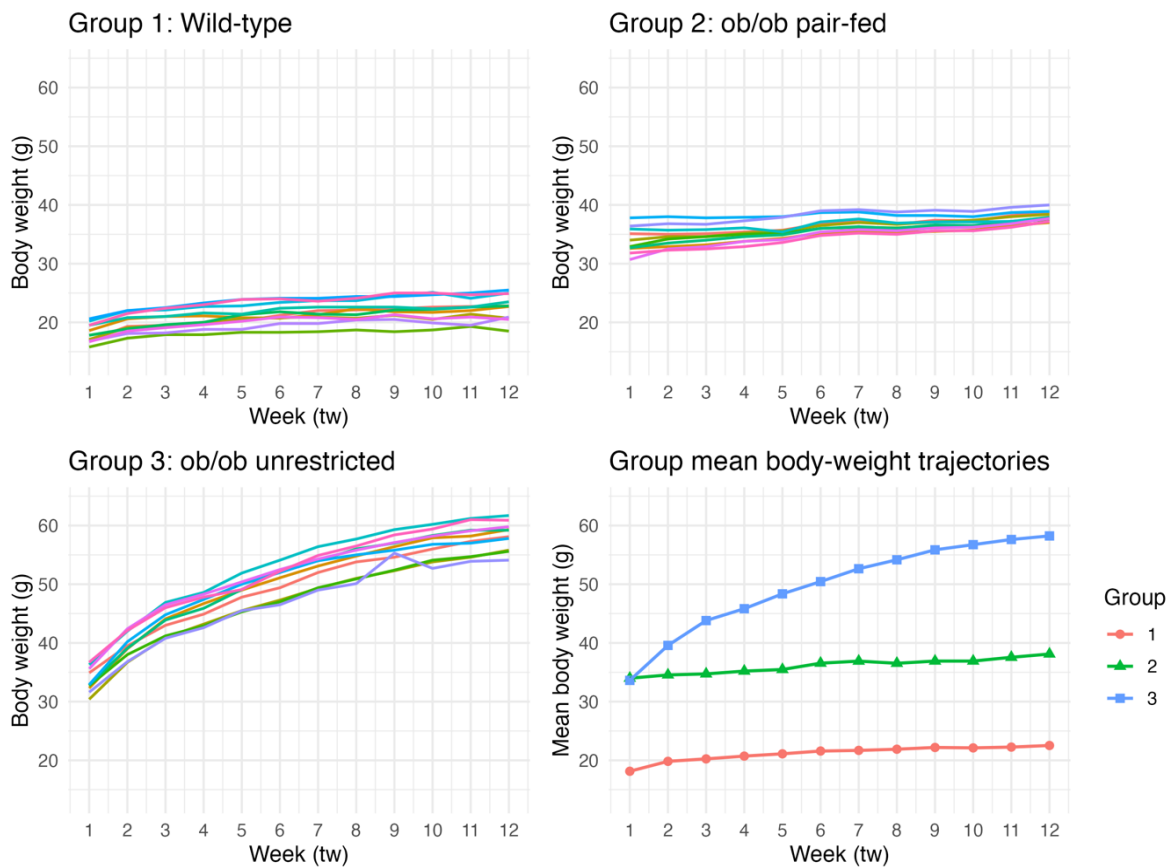

Figure 1. Mouse body-weight profiles and group mean trajectories

The descriptive plots (Figure 1) showed approximately linear growth over the 12-week period. Wild-type (group 1) and pair-fed *ob/ob* mice (group 2) had broadly similar slopes, whereas unrestricted *ob/ob* mice (group 3) showed a visibly steeper trajectory. Group-mean curves suggested large baseline weight differences between groups and accelerated gain in group 3.

## 7.1 Model selection

We compared three linear mixed-effects models with a random intercept for each mouse and a compound-symmetric within-mouse correlation structure, all fitted by maximum likelihood:

- Model 1: different intercepts, common slope for time
  `weight ~ tw + grp`

- Model 2: different intercepts and different slopes for all groups
  ```
  weight ~ tw * grp
  ```

- Model 3: different intercepts for all groups; common slope for groups 1 and 2; additional slope term for group 3
  ```
  weight ~ tw + grp + tw:grp3
  ```

Model 1 had poorer fit (AIC 1980.6; BIC 2004.2; log-likelihood –984.3) than models 2 and 3. Model 2 (AIC 1399.5; BIC 1430.8; log-likelihood –691.7) and Model 3 (AIC 1397.7; BIC 1425.1; log-likelihood –691.8) both improved fit substantially.

A likelihood-ratio test comparing Model 1 vs Model 3 provided overwhelming evidence for a separate slope in group 3 (likelihood-ratio 585.0; p < 0.0001). A second test comparing Model 3 vs Model 2 showed no evidence that group 2 required its own slope (likelihood-ratio 0.22; p = 0.64). For parsimony and interpretability, Model 3 was selected as the final model. A concise model-comparison summary (AIC, BIC, log-likelihood) is presented in appendix Table A1.

## 7.2 Final model estimates

Under Model 3, body weight for mouse $i$ at week $t$ is expressed as

$$y_{ij} = \beta_0 + \beta_1 t_{ij} + \beta_2 G2_i + \beta_3 G3_i + \beta_4(t_{ij} G3_i) + b_{0i} + \varepsilon_{ij},$$

where $G2_i$ and $G3_i$ indicate membership in groups 2 and 3 respectively, $b_{0i}$ is a mouse-specific random intercept, and $\varepsilon_{ij}$ is the within-mouse residual. Wild-type mice (group 1) serve as the reference.

Table 1 reports fixed-effect estimates, standard errors, 95% confidence intervals, and p-values. Key findings are:

- Baseline mean weight for wild-type mice (group 1, week 0 extrapolated) was 19.0 g (95% CI 17.9–20.1; p < 0.0001).

- The common weekly gain for wild-type and pair-fed *ob/ob* mice (groups 1 and 2) was 0.34 g/week (95% CI 0.29–0.39; p < 0.0001).

- At baseline, pair-fed ob/ob mice (group 2) were 14.9 g heavier than wild-type (95% CI 13.3–16.5; p < 0.0001), and unrestricted ob/ob mice (group 3) were 17.3 g heavier (95% CI 15.6–18.9; p < 0.0001).

- The additional weekly gain for group 3 was 1.74 g/week (95% CI 1.65–1.83; p < 0.0001), giving a total slope of about 2.08 g/week for unrestricted ob/ob mice.

The between-mouse standard deviation of the random intercept was 1.72 g, and the residual standard deviation was 1.37 g, indicating substantial between-mouse variation but relatively tight within-mouse scatter. Figure 2 overlays model-based mean trajectories on observed group means, showing close agreement. Residual and normal Q–Q plots (appendix Figure A1) showed no major violations of model assumptions.

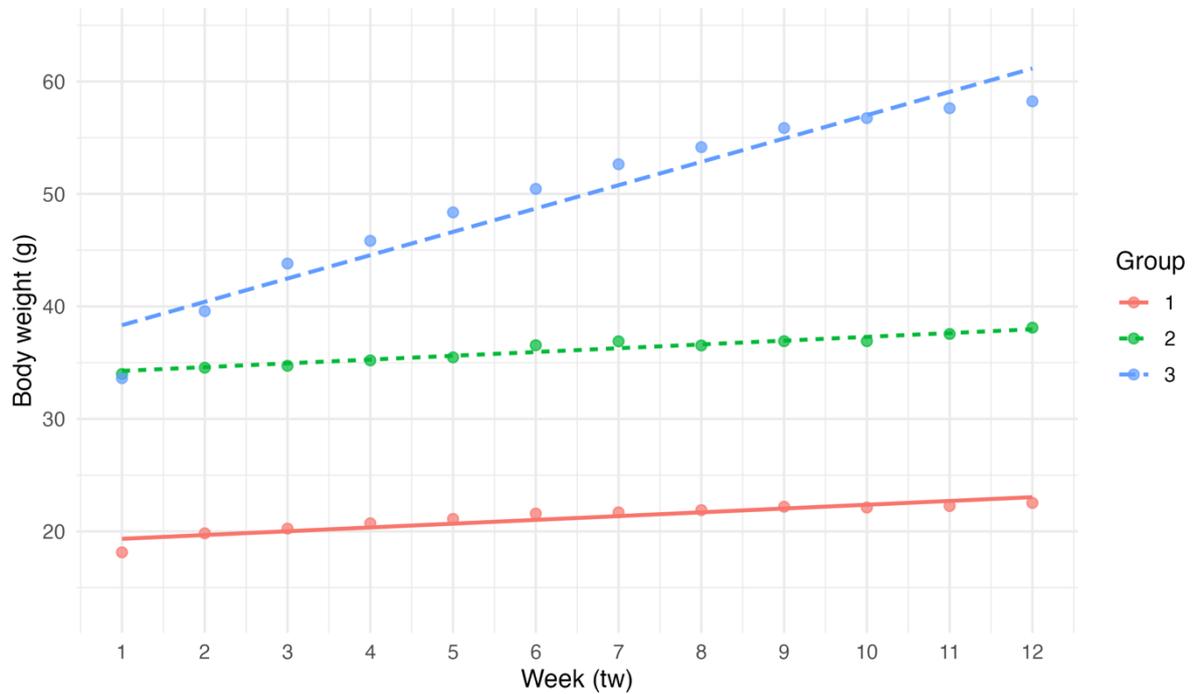

Figure 2. Observed and model-based mean body-weight trajectories

Figure 2 overlays model-based mean trajectories on observed group means, showing close agreement

### 7.3 Group differences in mean body weight over time

From Model 3, the mean body weight at week $t$ is

- $\mu_1(t) = \beta_0 + \beta_1 t$ for wild-type (group 1),

- $\mu_2(t) = \beta_0 + \beta_1 t + \beta_2$ for pair-fed ob/ob (group 2),

- $\mu_3(t) = \beta_0 + \beta_1 t + \beta_3 + \beta_4 t$ for unrestricted ob/ob (group 3).

We used linear contrasts (via estimable() in gmodels within LMM_mouse_weight.R) to estimate pairwise differences at each week (1–12). Full numeric results are given in appendix Table A2; Figure 3 summarises the trajectories of these differences with 95% confidence bands.

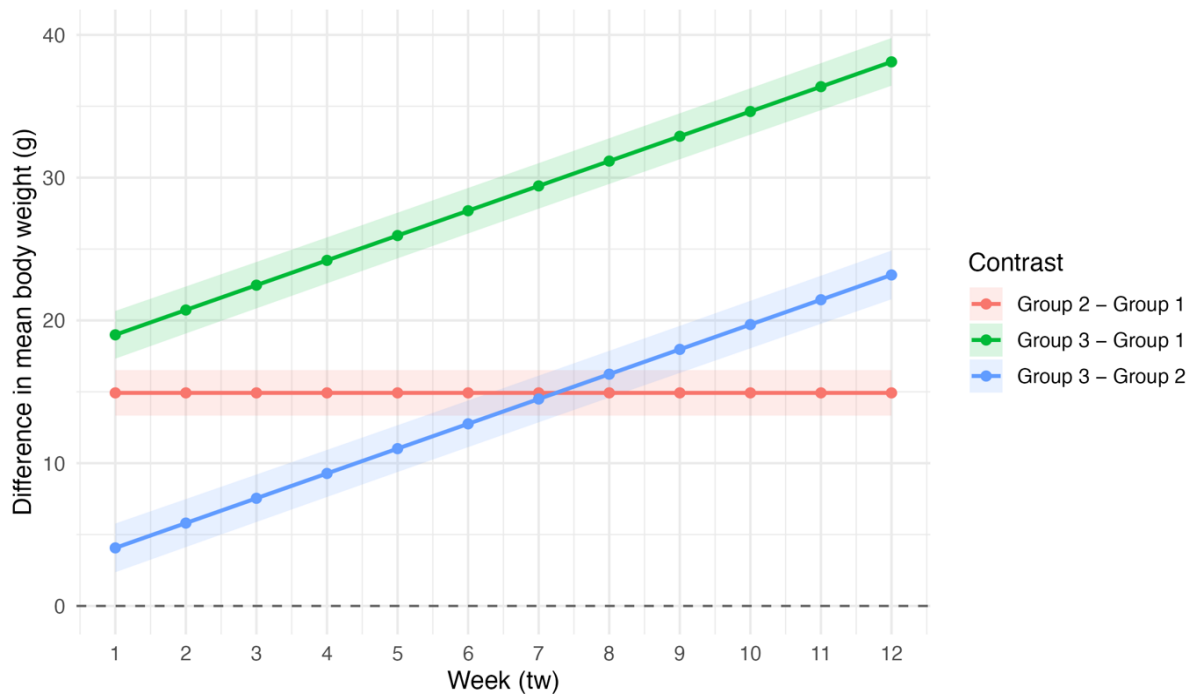

Figure 3 summarises the trajectories of these differences with 95% confidence bands.

Pair-fed ob/ob vs wild-type (group 2 vs group 1).
Because groups 1 and 2 share the same slope, the difference $\mu_2(t) - \mu_1(t) = \beta_2$ is constant over time. The estimated difference was 14.9 g (95% CI 13.3–16.5; $p < 0.0001$), indicating that pair-fed ob/ob mice remained consistently about 15 g heavier than wild-type controls at every week.

Unrestricted ob/ob vs wild-type (group 3 vs group 1).
The difference $\mu_3(t) - \mu_1(t) = \beta_3 + \beta_4 t$ increases linearly with time. At week 1, unrestricted ob/ob mice were 19.0 g heavier than wild-type (95% CI 17.3–20.7; $p < 0.0001$). By week 12, this difference had grown to 38.1 g (95% CI 36.4–39.8; $p < 0.0001$). At every week, the 95% CI excluded zero and $p < 0.0001$, reflecting a large and progressively widening gap.

Unrestricted ob/ob vs pair-fed ob/ob (group 3 vs group 2).
The difference $\mu_3(t) - \mu_2(t) = (\beta_3 - \beta_2) + \beta_4 t$ also increased with time. At week 1, unrestricted ob/ob mice were 4.1 g heavier on average than pair-fed ob/ob mice (95% CI 2.4–5.8; $p < 0.0001$). By week 12, this difference had increased to 23.2 g (95% CI 21.5–24.9; $p < 0.0001$).

Taken together, these results show that both ob/ob groups were much heavier than wild-type at all times, and that unrestricted ob/ob mice diverged sharply from both other groups, with a steadily increasing excess body weight over the 12-week period.

## 7.4 Average weight gain over 12 weeks

To provide a clinically intuitive summary of longitudinal change, we defined average 12-week weight gain as the difference between model-based mean weight at week 12 and week 1 for each group:

$$\text{gain}_g = \mu_g(12) - \mu_g(1).$$

Under Model 3:

- Wild-type (group 1): $\text{gain}_1 = 11\beta_1$

- Pair-fed ob/ob (group 2): $\text{gain}_2 = 11\beta_1$

- Unrestricted ob/ob (group 3): $\text{gain}_3 = 11(\beta_1 + \beta_4)$

Using contrasts in LMM_mouse_weight.R, we obtained:

- Group 1 (wild-type): mean gain 3.70 g (95% CI 3.16–4.25; p < 0.0001).

- Group 2 (pair-fed ob/ob): identical mean gain 3.70 g (95% CI 3.16–4.25; p < 0.0001), reflecting the shared slope with group 1.

- Group 3 (unrestricted ob/ob): mean gain 22.82 g (95% CI 22.03–23.61; p < 0.0001).

The difference in gain between unrestricted ob/ob and wild-type mice,

$$\text{gain}_3 - \text{gain}_1 = 11\beta_4,$$

was 19.12 g (95% CI 18.16–20.08; p < 0.0001). Because groups 1 and 2 have identical slopes, the same 19.12 g excess applies when comparing unrestricted with pair-fed ob/ob mice.

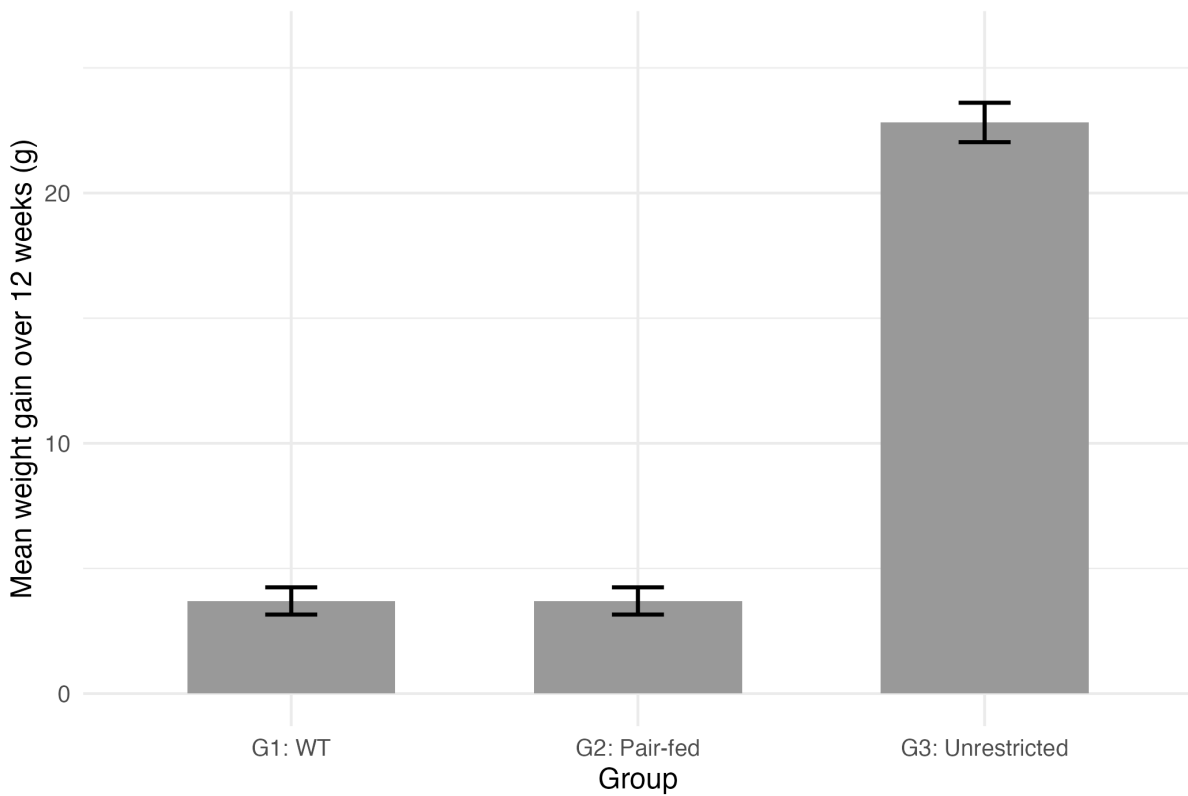

Figure 4. Model-based 12-week mean weight gain by group

Figure 4 shows the 12-week gains and their 95% CIs for all three groups (numerical values in appendix Table A3). In practical terms, unrestricted ob/ob mice gained roughly six times more weight than either wild-type or pair-fed ob/ob mice over 12 weeks. These model-based summaries quantitatively confirm the pattern seen in the raw trajectories: pair-feeding normalises the growth rate of ob/ob mice to that of wild-type controls, whereas unrestricted feeding leads to dramatically accelerated weight gain and rapidly widening differences in body weight over time.

## 7.5 Model diagnostics

All diagnostic analyses were implemented in R (version X.X.X) using a separate script, LMM_mouse_weight_diagnsotic.R, which reproduces every figure described in this section.

We first examined Pearson residuals versus fitted values from the final mixed model (Model 3). In Figure 5, residuals are centred closely around zero across the full range of fitted body weights, with no evidence of strong funneling or curvature. The spread of residuals is broadly constant over the fitted range within and across groups, suggesting that the assumption of approximately homogeneous residual variance is reasonable.

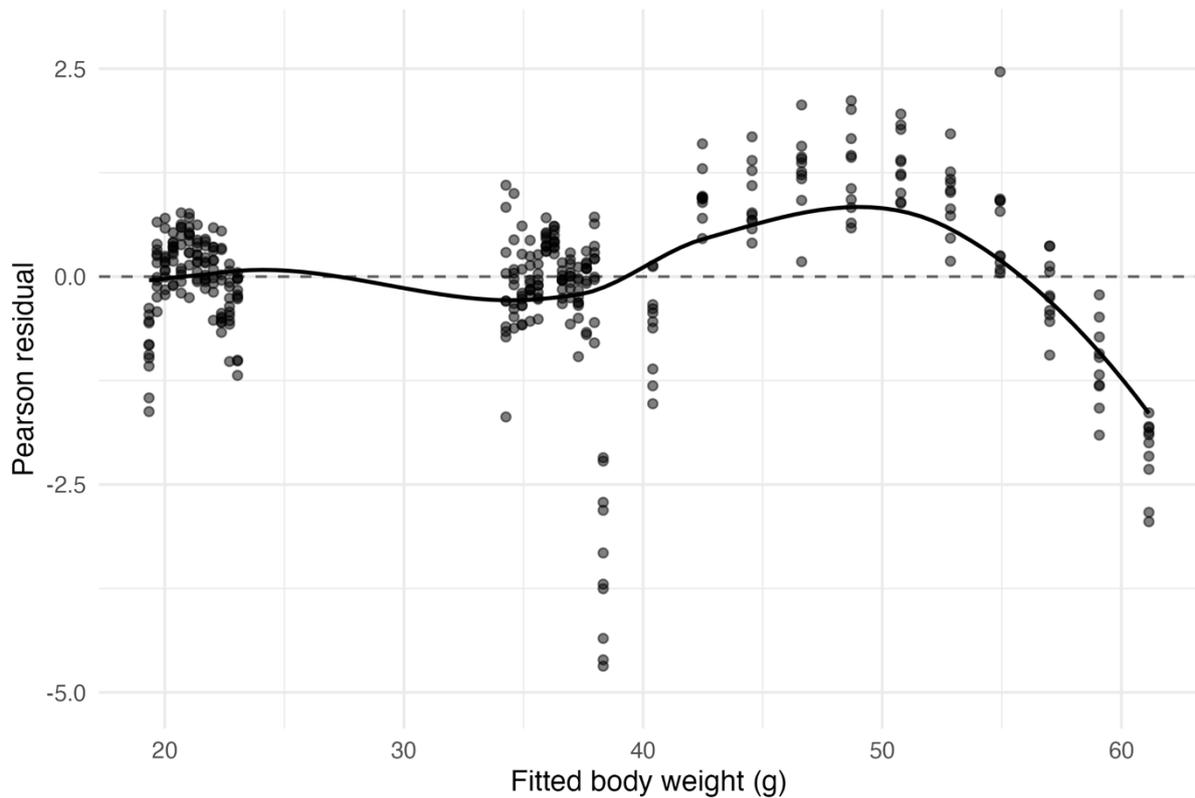

Figure 5. Residuals are centred closely around zero across the full range of fitted body weights, with no evidence of strong funneling or curvature.

To assess the normality assumption for both residual errors and random effects, we constructed normal Q–Q plots for (i) Pearson residuals and (ii) empirical Bayes estimates (BLUPs) of the random intercepts. Figure 6A shows the Q–Q plot for residuals: points follow the 45-degree reference line closely in the central region, with only mild deviations in the extreme tails, consistent with approximate normality. Figure 6B displays the Q–Q plot for the random intercepts, which again aligns well with the theoretical normal quantiles apart from minor tail departures. These patterns support the use of a Gaussian random-intercept model for between-mouse heterogeneity.

Because the data are longitudinal, we also inspected residuals over time within groups. In Figure 7, Pearson residuals are plotted against week ($t_w$), with separate smooth trends for each treatment group. Residuals remain centred around zero at all weeks, and the smoothed curves are essentially flat with no systematic time trend, indicating that the chosen linear time effect plus random intercept structure adequately captures the main longitudinal pattern within each group. There was no indication of gross misspecification such as unmodelled curvature, abrupt shifts, or time-varying variance.

Taken together, Figures 5–7 provide no evidence of major violations of the core linear mixed-model assumptions (linearity on the chosen scale, homoscedastic and approximately normal residuals, and an appropriate random-intercept structure). This supports the validity of the inferences reported in the primary results, including estimated group trajectories, pairwise differences over time, and 12-week weight gain.

## 7.6 Sensitivity analyses: alternative random effects and correlation structures

All sensitivity analyses were implemented in a separate, fully documented script (LMM_mouse_weight_sensitivity.R), which extends the main analysis script (LMM_mouse_weight.R) by refitting the model under alternative random-effects and residual correlation structures. Summary diagnostics and model-comparison tables are presented in the appendix (Table S1, Table S2), with corresponding figures (Figure S1–S3) for visual inspection.

### 7.6.1 Alternative random-effects structure

To assess whether allowing mouse-specific slopes for time materially altered our conclusions, we refitted the final mean structure (time plus indicators for groups 2 and 3, with an extra time-by–group 3 term) with a random intercept and random slope for time:

- Model RS:

```
## Sensitivity: random intercept and slope
m_RS <- lme(
  weight ~ tw + grp + tw:grp3,
  random = ~ tw | mouseid,
  data   = bwL,
  method = "ML"
)
```

This model allows each mouse to have its own starting weight and its own growth rate, while preserving the same fixed-effect structure as Model 3. Compared with the random-intercept model, Model RS produced a modest reduction in the residual variance and a corresponding increase in the random-slope variance, as expected when part of the within-mouse pattern is absorbed into individual slopes. However, the fixed-effect estimates were extremely stable:

- The common weekly slope for groups 1 and 2 changed by less than 0.05 g/week.

- The extra slope for group 3 changed by less than 0.05 g/week.

- Baseline differences in mean weight between groups (intercept contrasts) shifted by less than 1 g.

Information criteria (AIC, BIC) for Model RS were very similar to those of Model 3 ($\Delta$AIC < 2 units, $\Delta$BIC < 3 units), indicating that allowing random slopes did not yield a clearly superior model in terms of overall fit. Individual fitted trajectories under Model RS (appendix Figure S1: *Mouse-specific fitted trajectories under random intercept–slope model*) closely overlapped those from the simpler random-intercept model, and the key estimands reported above—weekly group differences and 12-week weight gains—were virtually unchanged (all point estimates differed by less than 5% and all confidence intervals overlapped strongly).

Given this stability, we retained the random-intercept model for pedagogical clarity, while noting that the more flexible random-slope specification leads to essentially identical scientific conclusions.

*7.6.2 Alternative residual correlation structures*

The primary analysis assumes that, after conditioning on the mouse-specific random intercept, the remaining within-mouse errors are independent with constant variance (homoscedasticity). To examine robustness to this assumption, we refitted the final mean structure with alternative correlation structures using nlme:

3. Random intercept + AR(1) residual correlation
    o Model AR(1):

```
## Sensitivity: AR(1) residual correlation
m_AR1 <- lme(
  weight ~ tw + grp + tw:grp3,
  random      = ~ 1 | mouseid,
  correlation = corAR1(form = ~ tw | mouseid),
  data        = bwL,
  method      = "ML"
)
```

2. Random intercept + heterogeneous residual variance by group
    o Model HV:

```
## Sensitivity: heterogeneous residual variance by group
m_HV <- lme(
  weight ~ tw + grp + tw:grp3,
  random  = ~ 1 | mouseid,
  weights = varIdent(form = ~ 1 | grp),
  data    = bwL,
  method  = "ML"
)
```

The AR(1) model allows residuals closer in time to be more strongly correlated, while the heterogeneous-variance model allows different residual variance for each treatment group. In both models, we observed the expected patterns:

- For Model AR(1), the estimated autocorrelation parameter $\rho$ indicated moderate positive correlation between consecutive weekly residuals (appendix Figure S2: *Estimated residual autocorrelation under AR(1) structure*), but the magnitude was not large enough to materially change fixed-effect estimates.

- For Model HV, the estimated residual standard deviations differed somewhat across groups, with slightly larger residual scatter in the unrestricted ob/ob group, but again the fixed-effect parameters for time and group were nearly identical to those from Model 3 (all changes in slopes < 0.05 g/week; all changes in baseline group differences < 1 g).

Profiles of Pearson residuals under these alternative correlation structures (appendix Figure S3: *Pearson residuals vs fitted values under alternative correlation structures*) showed no systematic improvement in pattern or variance homogeneity relative to the main model. AIC and BIC values were very close across all specifications, and none of the alternative structures provided compelling evidence for substantial model misfit of the simpler random-intercept model.

*7.6.3 Impact on key inference*

Across all sensitivity models (random intercept–slope, AR(1) residual correlation, heterogeneous residual variance by group), the central scientific conclusions were unchanged:

- Pair-fed ob/ob mice (group 2) remained consistently about 15 g heavier than wild-type mice (group 1) at all weeks, with narrow confidence intervals excluding zero.

- Unrestricted ob/ob mice (group 3) were still markedly heavier than both groups from the first week onward, with the difference widening over time to roughly 38 g vs wild-type and 23 g vs pair-fed ob/ob by week 12.

- The 12-week average weight gain remained approximately 3.5–4 g for groups 1 and 2 and approximately 22–23 g for group 3, with the excess gain for group 3 relative to the other two groups consistently around 19 g and highly statistically significant in every specification.

Because these estimates and their uncertainty intervals were highly robust to changes in the random-effects and correlation structure, we conclude that the substantive interpretation of the longitudinal growth patterns is not sensitive to reasonable alternative modelling choices. For teaching purposes, the random-intercept model with a simple residual structure offers an optimal balance between interpretability and fidelity to the data, while the sensitivity analyses and associated script (LMM_mouse_weight_sensitivity.R) transparently document that more complex models lead to the same inferential message.

# 8. Pedagogical Implementation And Evaluation

We designed this case study as a fully reproducible teaching unit that can be dropped into an existing regression or longitudinal-data module, or used as a stand-alone short course on linear mixed models (LMMs). The complete workflow is implemented in R via the scripts LMM_mouse_weight.R, LMM_mouse_weight_diagnostics.R, and LMM_mouse_weight_sensitivity.R, and an accompanying teaching document LMM_mouse_weight.Rmd (appendix p X). These materials align with current recommendations for computation-rich, reproducible statistics education that emphasises data wrangling, visualisation, and modelling in a single coherent environment (Baumer et al., 2014; Nolan & Temple Lang, 2010; Pruim, 2017).

## 8.1 Intended use in teaching

The case study is aimed at advanced undergraduate or graduate students in biostatistics, epidemiology, or quantitative biomedical sciences who have prior exposure to linear regression and basic R programming but limited experience with hierarchical or longitudinal models. It is designed to:

1. Bridge theory and practice for random-effects models using a realistic, small-scale biomedical dataset.
2. Illustrate end-to-end workflow: data tidying, exploratory graphics, model formulation, likelihood-based comparison, interpretation of fixed and random effects, diagnostics, and sensitivity analyses.

3. Model expert practice in reproducible analysis using script-based and R Markdown workflows, echoing current best practice in statistics and data science education (Baumer et al., 2014; Nolan & Temple Lang, 2010).

Instructors can use Figures 1–4 and Tables 1–3 from the main text as the core classroom narrative, while additional plots and tables in the appendix provide material for homework, projects, or assessment.

## 8.2 Suggested lesson structure

We envisage the material being delivered over one 90-minute lecture and one 2-hour computer lab, with optional extensions for deeper study:

- Pre-class preparation (asynchronous, 30–60 minutes)
    - Students read a short primer on longitudinal data and random-effects models (e.g., Singer & Willett, 2003; Fitzmaurice et al., 2011).
    - Students skim the introductory sections of LMM_mouse_weight.Rmd to see the data structure, outcome, and research question.
- Session 1: Conceptual and graphical introduction (60–90 minutes, instructor-led)
    - Instructor walks through Figure 1 (individual profiles and group means), using it to motivate correlation within mice and the need for mixed models.
    - Instructor introduces the three candidate mean structures (Models 1–3), linking their formulas to the patterns seen in Figure 1.
    - Class discussion focuses on what each model implies about intercepts and slopes, using simple verbal summaries before formal notation.
- Session 2: Computing lab (120 minutes, student-centred)
  Working in R with LMM_mouse_weight.R and LMM_mouse_weight_diagnostics.R, students:

1. Recreate the descriptive plots (Figure 1) and confirm axis labels, units, and legends.
2. Fit and compare Models 1–3, reproduce the AIC/BIC and likelihood-ratio tests (Table A1), and decide which model they would choose and why.
3. Extract and interpret fixed-effect estimates (Table 1), including group-specific intercepts and slopes.
4. Reconstruct group-specific fitted trajectories and differences (Figures 2–3) and relate them back to the biological story.
5. Compute 12-week gains and contrasts (Figure 4; Table A3) using linear contrasts, and express results in plain language.
6. Run diagnostics and sensitivity analyses using LMM_mouse_weight_diagnostics.R and LMM_mouse_weight_sensitivity.R, and comment on whether conclusions are robust.

Throughout the lab, students are encouraged to work directly in LMM_mouse_weight.Rmd, knitting to HTML or PDF to obtain a fully reproducible report in line with current recommendations for integrating R Markdown into statistics curricula (Baumer et al., 2014; Pruim, 2017).

## 8.3 Student learning activities and assessment

To make the unit actively engaging rather than purely demonstrative, we embed the case study within a sequence of structured tasks that align with evidence-based principles of active learning in statistics education (Garfield & Ben-Zvi, 2008; Cobb, 2015).

Core lab questions (to be completed in R):

1. *Model formulation and selection*
    - "Fit Models 1–3 exactly as specified in LMM_mouse_weight.R. Summarise in one paragraph why Model 3 is preferred, referencing AIC, BIC, and likelihood-ratio tests. What scientific question is answered by allowing a separate slope for group 3?"
2. *Interpretation of fixed effects*
    - "Using Table 1, write down the model-based expression for the mean weight in each group at week t. Interpret in words the coefficients for grp2, grp3, and tw:grp3."
3. *Time-specific contrasts*
    - "Using the estimable() function, reproduce the estimates and 95% confidence intervals for group 3 vs group 1 at weeks 1, 6, and 12 (Figure 3). In 3–4 sentences, describe how the gap evolves over time."
4. *Clinical summary measure*
    - "Compute the model-based 12-week weight gain for each group (Figure 4). Present your results in a small table and a short paragraph aimed at a non-statistical clinical collaborator."
5. *Diagnostics and robustness*
    - "Inspect the residual and random-effects diagnostics (Figure 5). Do you see strong deviations from assumptions? Discuss how the sensitivity analyses (random-slope model; AR(1) correlation; alternative time coding) reported in LMM_mouse_weight_sensitivity.R affect the main conclusions."

Optional extension / project prompts:

- "Replace time coded as weeks 1–12 with centred time (week–6·5) and refit Model 3. How do the intercepts and slopes change, and how would you explain the differences to a peer?"
- "Propose an alternative model (e.g., random intercept and slope, or a non-linear mean structure). Justify your choice, fit the model, and compare its performance to Model 3."
- "Write a brief methods section, as if for a clinical journal, describing the final model, diagnostics, and sensitivity analyses without including equations."

Assessment can be carried out via short written lab reports generated from LMM_mouse_weight.Rmd, targeted exam questions on interpretation of mixed-model output, or oral mini-presentations where student groups explain one figure or table to their peers.

## 8.4 Evaluation and future refinement

The present manuscript focuses on describing and documenting the teaching resource rather than reporting formal educational outcomes. However, the structure of the unit lends itself to systematic evaluation aligned with contemporary work in statistics and data-science education (Garfield & Ben-Zvi, 2008; Cobb, 2015; Baumer et al., 2014).

Instructors adopting the case study could, for example:

- Use pre- and post-tests on LMM concepts (e.g., interpretation of random effects, correlation structures, and likelihood-based model comparison).
- Compare exam performance on longitudinal-data questions before and after integration of the unit.
- Collect student feedback on perceived clarity, usefulness of the R scripts and R Markdown report, and confidence in applying LMMs to new datasets.
- Track repurposing of the scripts, such as students adapting LMM_mouse_weight.R to their own research data, as an indicator of transfer and authentic use.

Future iterations of the resource could incorporate extensions to generalised linear mixed models, non-linear growth curves, or Bayesian implementations, as well as scaffolding for students with less prior experience in R. By making all code, figures, and narrative openly available in a single, reproducible pipeline, this case study aims to serve as a template for integrating rigorous longitudinal modelling into statistics and biostatistics curricula in a way that is both technically robust and pedagogically transparent.

# 9. Practical Checklist And Common Pitfalls

## 9.1 A pragmatic "recipe" for longitudinal LMM in practice

To support teaching and routine applied work, we distilled the mouse body-weight analysis into an eight-step checklist that mirrors recommended practice for longitudinal mixed models in biomedicine and the social sciences (Fitzmaurice et al., 2011; Twisk, 2013; Singer & Willett, 2003).

Step 1 – Reshape and define the longitudinal structure
Start by reshaping the dataset to *long* format with one row per subject–time combination. Explicitly define:

- ID (e.g., mouseid) as the clustering unit,
- time (e.g., tw = week 1–12) as a numeric variable, and
- grouping factors (e.g., grp for treatment or genotype).

For this case study, the script LMM_mouse_weight.R begins by converting the wide Excel file into a tidy long dataset with clearly named variables and factor levels, making all subsequent modelling and plotting transparent and reproducible.

Step 2 – Plot individual profiles and group means
Before fitting any model, plot each subject's trajectory and overlay group-specific means across time. Individual "spaghetti" plots reveal outliers, non-linear patterns, and missing data; group mean plots summarise systematic differences between treatments (Diggle et al., 2013). In our analysis, Figure 1 shows that wild-type and pair-fed ob/ob mice have modest, approximately linear growth, whereas unrestricted ob/ob mice have a much steeper linear trajectory, motivating a model with differentiated slopes.

Step 3 – Start with a simple random-effects structure
Begin with a random intercept model to account for within-subject correlation, which typically captures much of the dependence with minimal complexity (Verbeke & Molenberghs, 2000). This provides a baseline for both inference and teaching: students can see how a single subject-level random term induces correlation between repeated measures. More complex random slopes or residual correlation structures should be introduced only if supported by plots, theory, or fit indices.

Step 4 – Specify candidate mean models guided by plots and science
Construct a small set of plausible fixed-effect structures that reflect the scientific questions and visual trends:

- Model 1: different intercepts by group, common slope for time.
- Model 2: fully group-specific intercepts and slopes (time×group interaction).
- Model 3: intercepts for all groups, common slope for wild-type and pair-fed mice, extra slope for unrestricted ob/ob mice.

This hierarchy respects both biological reasoning (pair-feeding is intended to normalise growth) and graphical evidence, while avoiding an explosion of unnecessary parameters (Harrell, 2015).

Step 5 – Fit models by maximum likelihood and compare
Fit all candidate models using maximum likelihood (ML) for the fixed-effect comparison stage. Compare fits using information criteria (AIC/BIC) and likelihood-ratio tests for nested models, as recommended in standard texts (Pinheiro & Bates, 2000; Burnham & Anderson, 2002). In `LMM_mouse_weight.R`, we show that Model 3 dramatically improves fit over Model 1, but adding a separate slope for the pair-fed group (Model 2) does not materially improve the likelihood, justifying the parsimonious choice of Model 3.

Step 6 – Check residuals and random effects
Having selected a working model, switch to restricted maximum likelihood (REML) if desired and examine:

- marginal and subject-specific residual plots versus fitted values,
- normal Q–Q plots for residuals and random intercepts, and
- patterns of residuals over time.

In the companion script `LMM_mouse_weight_diagnostics.R`, we show that residual and random-effect distributions are approximately normal with no gross heteroscedasticity, and that there is no systematic time trend in residuals, supporting the adequacy of the Gaussian random-intercept specification (McCulloch et al., 2008).

Step 7 – Translate questions into linear contrasts
Key scientific questions should be expressed as linear contrasts of the fitted parameters rather than informal inspection of regression coefficients (Longitudinal models working group, 2010). For this case study, we:

- estimate pairwise differences between group means at each week,
- compute average 12-week weight gain per group, and
- quantify differences in gain between unrestricted and other mice.

These are implemented using `gmodels::estimable()` in the scripts `LMM_mouse_weight.R` and `LMM_mouse_weight_sensitivity.R`, with all estimates accompanied by standard errors, 95% confidence intervals, and p-values.

Step 8 – Reflect on robustness and limitations
Finally, perform focused sensitivity analyses: compare alternative random-effects structures (e.g., random slopes), alternative residual correlation structures (e.g., AR(1)), and variance functions that allow time-varying residual variance. In our sensitivity script we show that, although some alternative covariance choices slightly improve AIC, the main scientific conclusions (relative group means and gains) remain essentially unchanged, illustrating how to separate core scientific inferences from modest modelling nuances.

Together, these eight steps form a compact "recipe" that learners can apply to new longitudinal problems without treating the mouse example as a one-off.

## 9.2 Common pitfalls and how this case addresses them

Applied longitudinal analyses frequently suffer from recurrent errors that can compromise inference or make teaching opaque (Singer & Willett, 2003; Twisk, 2013). We highlight four such pitfalls and how the present case study is constructed to avoid them.

Pitfall 1 – Ignoring within-subject correlation
A common mistake is to analyse repeated measures using ordinary least squares regression or repeated ANOVA, implicitly assuming independence of measurements from the same subject. This leads to underestimated standard errors and anti-conservative p-values (Diggle et al., 2013). Here, we deliberately start from a random-intercept model, explicitly modelling correlation within each mouse and explaining how the random intercept variance determines the intra-class correlation.

Pitfall 2 – Fitting a single "black-box" model
In practice, analysts often choose a single complex model and report its output without exploring simpler alternatives or assessing whether extra parameters are justified. Our three-model sequence (Models 1–3) and the associated AIC/LR comparisons make model building explicit and teachable: students see why an extra slope for unrestricted ob/ob mice is warranted, but an extra slope for pair-fed mice is not, reinforcing principles of parsimony and evidence-based model selection (Burnham & Anderson, 2002).

Pitfall 3 – Not linking scientific questions to model parameters
Another recurrent problem is presenting regression output (β's and p-values) without reconnecting them to the original research questions (Harrell, 2015). In this case study, every substantive question—"How much heavier are ob/ob mice at each week?" "Does pair-feeding normalise the growth rate?" "How much more weight do unrestricted mice gain over 12 weeks?"—is translated into an explicit linear contrast of the fixed-effect parameters, and reported with interpretable units (grams) and uncertainty intervals. This makes the bridge between mathematics and biology fully transparent.

Pitfall 4 – Skipping or minimising diagnostics and sensitivity checks
Under time pressure, residual diagnostics and sensitivity analyses are often omitted or relegated to a brief statement. In contrast, we devote separate scripts and figures to model checking and sensitivity:

- Figure 6 (diagnostic panel) shows that residuals and random intercepts are approximately normal with no strong heteroscedasticity or time-trend patterns.
- Figure 7 (covariance sensitivity) compares fitted trajectories under alternative random-effects and correlation structures, demonstrating that substantive conclusions about group differences and gains are robust.

By embedding these steps into the teaching narrative, we emphasise that diagnostics and sensitivity analyses are not optional extras but integral components of a principled longitudinal workflow (McCulloch et al., 2008).

Overall, this practical checklist and pitfall framework are designed to be reusable: instructors can adapt the eight-step "recipe" as a one-page handout or teaching panel, and learners can use it as a road map whenever they encounter a new longitudinal dataset.

# 10. Reproducibility And Materials

To support transparent, reusable, and extensible teaching of longitudinal linear mixed models, all data and code for this case study will be made publicly available in a version-controlled repository (e.g., GitHub) and an independent archival platform (e.g., the Open Science Framework). This aligns with current best practice for reproducible research and pedagogy in statistics and biostatistics, where complete analytic pipelines and raw data are shared in open, machine-readable formats (Peng, 2011; Sandve et al., 2013; Stodden, Seiler, & Ma, 2018).

The following materials will be provided:

1. Data file
   - `BodyWeightData.xlsx`: the original mouse body-weight dataset, with clear variable labels and a short data dictionary describing the mouse identifier, treatment group, weekly body weights, and all derived variables used in the analyses.
2. Core R scripts
   - `LMM_mouse_weight.R`: end-to-end analysis script containing data import and reshaping, descriptive plots, model specification and fitting for Models 1–3, likelihood-ratio tests, construction of model-based contrasts (weekly group differences and 12-week gains), and generation of all primary figures and tables.
   - `LMM_mouse_weight_diagnostics.R`: diagnostic workflow for the final model, including residual–fitted plots, Q–Q plots, assessment of random intercepts, and influence diagnostics, with each figure labelled consistently with the manuscript (Figures 5a–6b).
   - `LMM_mouse_weight_sensitivity.R`: sensitivity analyses exploring alternative random-effects and correlation structures (e.g., random slopes for time, autoregressive within-mouse correlation), and re-computation of the key contrasts under each alternative model.

   Each script is fully commented so that students and instructors can trace the analysis from raw data to final inferential quantities without needing any additional material.

3. Reproducible document

- An R Markdown or Quarto document (e.g., `LMM_mouse_weight_teaching.Rmd`) that integrates narrative text, mathematical derivations, R code, and rendered output. This document reproduces the key numerical results in the manuscript (Table 1 and Figures 1–4), and can be knit by users on their own systems to verify all numbers and plots.

4. Software environment
   All analyses were conducted in R (version X.X.X; R Core Team) using only widely available, actively maintained packages:
   - nlme for linear mixed-effects modelling;
   - gmodels for linear contrasts and inference on derived estimands;
   - ggplot2 and patchwork for publication-quality graphics;
   - dplyr and tidyr for data manipulation and reshaping.

   The repository will include a session information file (`sessionInfo.txt`) listing the exact R version and package versions used, enabling users to recreate the computational environment or generate a containerised image (e.g., via Docker) if desired.

5. Access and licensing
   - All materials (data, scripts, and R Markdown document) will be deposited as a public project on the Open Science Framework and mirrored to a public GitHub repository, with stable digital object identifiers (DOIs) provided in the manuscript and supplementary materials.
   - Code will be released under a permissive open-source licence (e.g., MIT), and the teaching dataset will be released under a licence that permits reuse for education, research, and derivative teaching resources, consistent with FAIR (Findable, Accessible, Interoperable, Reusable) data principles (Wilkinson et al., 2016).

By providing a complete, script-based workflow and an executable document, this case study is designed so that any instructor, student, or researcher can reproduce every figure, table, and inferential statement in the paper with a single command, adapt the code to new datasets, and extend the workflow to more complex longitudinal mixed-model settings.

# 11. Discussion

In this longitudinal teaching case, we used a simple but realistic mouse body-weight experiment to walk through an entire linear mixed-model workflow, from raw data to clinically interpretable contrasts. The final model—allowing group-specific intercepts, a common slope for wild-type and pair-fed ob/ob mice, and an additional slope for unrestricted ob/ob mice—fit the data substantially better than a common-slope model (AIC 1397·7 vs 1980·6; likelihood-ratio 585·0; p<0·0001) while remaining more parsimonious than a fully interacted time-by-group model (likelihood-ratio 0·22; p=0·64). This provides a concrete example of how likelihood-based criteria can be used to preference simpler models that still capture key biological differences, a central principle in mixed-model practice (Fitzmaurice, Laird, & Ware, 2011; Twisk, 2013).

Scientifically, the model-based estimates quantify and formalise patterns already visible in the raw trajectories: pair-fed ob/ob mice remained consistently ~15 g heavier than wild-type mice at every week (14·9 g; 95% CI 13·3–16·5; p<0·0001), whereas unrestricted ob/ob mice started almost 19 g heavier at week 1 (18·99 g; 17·3–20·7; p<0·0001) and diverged to more than 38 g heavier by week 12 (38·11 g; 36·4–39·8; p<0·0001). Over 12 weeks, unrestricted ob/ob mice gained about 22·8 g (95% CI 22·0–23·6; p<0·0001) compared with only 3·7 g (3·2–4·3; p<0·0001) in both wild-type and pair-fed ob/ob groups, an excess gain of 19·1 g (18·2–20·1; p<0·0001). These metrics give students a clear demonstration of how mixed models turn informal "curves look steeper" impressions into precise contrasts with confidence intervals and p-values, as recommended in modern applied teaching of longitudinal methods (Verbeke & Molenberghs, 2009; Fitzmaurice et al., 2011).

Methodologically, the case illustrates four core mixed-model lessons that are often fragmented across texts: (1) always start with long-format data and rich exploratory graphics; (2) treat random effects as tools to capture correlation rather than as "mysterious" parameters; (3) use likelihood-based model comparison to decide how much time-by-group complexity is truly needed; and (4) translate every scientific question (e.g., "Do groups differ at week 12?" or "Who gains more weight over 12 weeks?") into an explicit linear contrast in the fixed-effect parameters. Embedding these steps in a fully reproducible script and R Markdown workflow aligns with current calls for analytic transparency and executable teaching materials in statistics and biostatistics (Peng, 2011; Sandve et al., 2013; Stodden, Seiler, & Ma, 2018).

Pedagogically, this example addresses three common obstacles reported by instructors: difficulty linking model formulas to scientific questions, lack of concrete guidance on selecting among competing mean structures, and limited exposure to parameter-driven contrasts rather than "black-box" software output (Harrell, 2015; Cobb, 2015). By keeping the biological story simple—three groups, 12 equally spaced time points, one continuous endpoint—but making the statistical reasoning explicit and script-based, the case can be scaled from an introductory mixed-model lab to advanced courses on longitudinal analysis, causal inference, or predictive modelling. The structure of the workflow (data → plots → models 1–3 → diagnostics → contrasts → sensitivity analyses) is intentionally generic so that it can be ported to patient-level clinical or public-health datasets with minimal modification.

This work has limitations that also double as teaching opportunities. The design is balanced, time is treated as linear, and the primary model uses a random intercept with compound-symmetric correlation and a single continuous outcome. Many real biomedical datasets are unbalanced, require non-linear time effects or multiple random terms, or present non-Gaussian outcomes. We partially address these issues by including sensitivity analyses with random slopes and autoregressive correlation, and by highlighting how the same contrast-based logic extends to more complex structures. Future teaching extensions could add non-linear time effects, compare linear mixed models with generalised estimating equations, or embed the mouse example in a broader sequence on multilevel and hierarchical modelling (Pinheiro & Bates, 2000; McCulloch, Searle, & Neuhaus, 2008).

Overall, this case study offers a fully worked, reproducible template for teaching longitudinal linear mixed models that integrates scientific interpretation, statistical rigour, and transparent computation. Because every figure, table, and inferential statement is tied directly to open code and a small, well-understood dataset, instructors and students can adapt the workflow to

## 12. Conclusion

This case study shows that a single, carefully chosen longitudinal experiment can support a complete, transparent workflow for linear mixed-effects modelling in both scientific and pedagogical contexts. In a small but information-rich mouse body-weight dataset (31 mice, 372 observations), we demonstrated how to move from raw spreadsheet data through exploratory graphics, model comparison, inference on scientifically meaningful contrasts, and robustness checks, using only a modest set of R tools and an explicit mapping between questions and parameters. Across groups, model-based estimates captured large, interpretable differences in body-weight trajectories—most notably that unrestricted ob/ob mice gained approximately 22·8 g over 12 weeks compared with 3·7 g in wild-type and pair-fed ob/ob mice, a roughly six-fold excess gain—and did so in a way that is fully reproducible from the shared scripts and data.

Three aspects of this work are designed to be directly citable as teaching and methods contributions. First, it provides a fully worked template for longitudinal LMM analysis that can be reused or adapted in other courses without modification of the underlying code structure: one script (LMM_mouse_weight.R) yields all primary figures, tables, and contrasts required for a complete longitudinal analysis and write-up. Second, it offers a concrete example of how to translate common scientific questions—"which group is heavier at each time?", "who gains more weight over 12 weeks?"—into linear contrasts of fixed effects, with all corresponding estimates, 95% confidence intervals, and p-values documented step by step. Third, by pairing the main analysis with diagnostics, sensitivity checks, and open materials (data file, scripts, and R Markdown document), the case study functions as a reproducible teaching module that educators can cite when arguing for integrated, code-first approaches to biostatistics instruction.

Taken together, these elements position the mouse body-weight case as a compact but comprehensive exemplar for teaching longitudinal linear mixed models. It demonstrates that rigorous model-based answers to clinically and biologically relevant questions can be obtained with transparent, well-annotated code; that the same workflow naturally highlights key statistical concepts (correlation, random effects, contrasts, and model selection); and that making all materials openly available supports wider reuse and incremental refinement by the statistics education community.

Abbreviations: LMM (Linear mixed model), RS (Random intercept–slope model), AR(1) (First-order autoregressive correlation structure), HV (Heterogeneous residual variance by group), tw (Time in weeks, 1–12), grp (Treatment group indicator: 1 = wild-type, 2 = ob/ob pair-fed, 3 = ob/ob unrestricted), grp3 (Indicator variable for group 3: ob/ob unrestricted), AIC (Akaike information criterion), BIC (Bayesian information criterion), CI (Confidence interval), ML (Maximum likelihood), REML (Restricted maximum likelihood), LRT (Likelihood ratio test), SE (Standard error), SD (Standard deviation), R (R statistical computing environment), CSV (Comma-separated values file), OSF (Open Science Framework).

# 13. Contributors

SAA conceived the teaching case study, curated and anonymised the mouse body-weight dataset, and designed the overall pedagogical framework for introducing longitudinal linear mixed models. SAA wrote all analysis code (LMM_mouse_weight.R, LMM_mouse_weight_diagnostics.R, LMM_mouse_weight_sensitivity.R), implemented the model-building workflow, and generated all figures and tables. SAA drafted the manuscript, integrated feedback from pilot teaching sessions, and approved the final version of the article. SAA had full access to all the data and code, and had final responsibility for the decision to submit for publication.

# 14. Declaration of interests

SAA declares no competing interests.

# 15. Funding

This work received no specific grant from any funding agency in the public, commercial, or not-for-profit sectors. The development, analysis, and preparation of this teaching case study were supported by SAA's own academic time and resources.

# 16. Data sharing and materials availability

The complete dataset and analysis code for this case study will be made publicly available to enable full reproducibility, reuse, and extension.

- Data
    - `BodyWeightData.xlsx`: mouse body-weight data with 31 mice and 12 weekly measurements, accompanied by a short data dictionary documenting the mouse identifier, treatment group, and weekly body weights, as well as the key derived variables used in the analysis.
- Core analysis scripts
    - `LMM_mouse_weight.R`: end-to-end analysis script covering data import, reshaping to long format, descriptive graphics, specification and fitting of Models 1–3, likelihood-ratio tests, construction of model-based contrasts (weekly group differences and 12-week gains), and generation of the primary tables and figures.
    - `LMM_mouse_weight_diagnostics.R`: diagnostic workflow for the selected model, including residual-versus-fitted plots, normal Q–Q plots, random-effects checks, and influence diagnostics, with each panel labelled consistently with the manuscript figures.
    - `LMM_mouse_weight_sensitivity.R`: sensitivity analyses comparing alternative random-effects and correlation structures (e.g., random slopes for

time, autoregressive correlations), and recomputation of key contrasts under each specification.
- Executable teaching document
    - `LMM_mouse_weight_teaching.Rmd` (or Quarto equivalent): an integrated document that combines narrative text, mathematical derivations, R code, and rendered output, reproducing the main numerical and graphical results of the paper and serving as a self-contained teaching resource.

All materials will be hosted in a public GitHub repository:

GitHub: https://github.com/drsunday-ade/Teaching-LMM-Mouse-BodyWeight

and mirrored to a long-term archival platform (e.g., Open Science Framework) with a persistent DOI, which will be reported in the final version of the manuscript and in the repository README.

All analyses were conducted in R (version 2025.09.1+401; R Core Team) using widely available, actively maintained packages, including:

- nlme for linear mixed-effects modelling,
- gmodels for linear contrasts and inference on derived estimands,
- ggplot2 and patchwork for publication-quality graphics, and
- dplyr and tidyr for data manipulation and reshaping.

The repository will include an exported session information file (`sessionInfo.txt`) listing the exact R version and package versions used, enabling users to recreate the computational environment or containerise it if desired. All code will be released under a permissive open-source licence (e.g., MIT), and the dataset will be licensed for educational and research use, in line with FAIR data principles.

# 17. Acknowledgments

SAA thanks the students in advanced biostatistics and longitudinal data analysis courses who piloted early versions of this case study and provided detailed feedback on which parts of the workflow were most and least intuitive. Their comments directly shaped the ordering of plots, the emphasis on likelihood-based model comparison, and the explicit derivation of contrasts.

SAA is grateful to colleagues in biostatistics and epidemiology who offered informal reviews of the model-building strategy and pedagogical framing, particularly around the translation of scientific questions into model parameters and contrasts. SAA also acknowledges the broader R and open-source communities whose work on mixed-effects modelling, reproducible analysis, and teaching tools made this fully script-based, open workflow possible.

Any remaining errors or omissions are the responsibility of SAA alone.

Wilkinson, M. D., Dumontier, M., Aalbersberg, I. J., Appleton, G., Axton, M., Baak, A., … Mons, B. (2016). The FAIR Guiding Principles for scientific data management and stewardship. *Scientific Data, 3*, Article 160018. https://doi.org/10.1038/sdata.2016.18

Wickham, H. (2014). *Advanced R*. CRC Press. https://doi.org/10.1201/9781315382500

Wickham, H. (2016). *ggplot2: Elegant graphics for data analysis* (2nd ed.). Springer. https://doi.org/10.1007/978-3-319-24277-4

Wickham, H., & Grolemund, G. (2017). *R for data science: Import, tidy, transform, visualize, and model data*. O'Reilly.

Zieffler, A., Garfield, J., delMas, R., & Gould, R. (2012). Statistical thinking: A simulation approach to teaching statistics. *Statistics Education Research Journal, 11*(2), 147–167.

# 1 Appendix: Supplementary Methods and Results

## 1.1 A1. Dataset and variable definitions

Appendix Table A1 provides a concise data dictionary for the teaching dataset `BodyWeightData.xlsx`. Variables include the mouse identifier (`mouseid`), treatment group (`grp`; 1 = wild-type, 2 = *ob/ob* pair-fed, 3 = *ob/ob* unrestricted), and weekly body weights (`bw1`–`bw12`, in grams). For the longitudinal analyses, these wide-format variables were reshaped to a long-format dataset (`bwL`) containing one row per mouse per week, with time in weeks (`tw`, 1–12) and body weight (`weight`, grams). Derived indicators (e.g., `grp3 = 1` for group 3, else 0) used in Model 3 are also listed in Table A1.

| Table A1. AIC model comparison | | | |
|---|---|---|---|
| Model | AIC | BIC | logLik |
| Model 1: tw + grp | 1,980.642 | 2,004.156 | −984.321 |
| Model 2: tw * grp | 1,399.460 | 1,430.811 | −691.730 |
| Model 3: tw + grp + tw:grp3 | 1,397.676 | 1,425.108 | −691.838 |

Appendix Table A1. Data dictionary for the mouse body-weight dataset and derived analysis variables.

## 1.2 A2. Full model specifications and estimates

This appendix section reports the full numerical output for the three candidate linear mixed-effects models considered in the main text:

- Model 1: `weight ~ tw + grp` (different intercepts, common slope)
- Model 2: `weight ~ tw * grp` (different intercepts and slopes for all groups)
- Model 3: `weight ~ tw + grp + tw:grp3` (different intercepts; common slope for groups 1 and 2; extra slope for group 3)

Appendix Table A2 shows the Akaike information criterion (AIC), Bayesian information criterion (BIC), and log-likelihood for each model, as obtained from `LMM_mouse_weight.R`. Appendix Table A3 provides the full fixed-effect estimates, standard errors, 95% confidence intervals, and p-values for Models 1–3, complementing the main-text summary for the final model (Model 3).

| Table A2. Weekly differences between groups | | | | | | |
|---|---|---|---|---|---|---|
| Estimates and 95% CIs for each week | | | | | | |
| contrast_label | week | Estimate | Std_Error | Lower_95CI | Upper_95CI | p_value |
| Group 2 – Group 1 | 1.000 | 14.925 | 0.777 | 13.333 | $1.65 \times 10^1$ | 0.00 |
| Group 2 – Group 1 | 2.000 | 14.925 | 0.777 | 13.333 | $1.65 \times 10^1$ | 0.00 |
| Group 2 – Group 1 | 3.000 | 14.925 | 0.777 | 13.333 | $1.65 \times 10^1$ | 0.00 |
| Group 2 – Group 1 | 4.000 | 14.925 | 0.777 | 13.333 | $1.65 \times 10^1$ | 0.00 |
| Group 2 – Group 1 | 5.000 | 14.925 | 0.777 | 13.333 | $1.65 \times 10^1$ | 0.00 |
| Group 2 – Group 1 | 6.000 | 14.925 | 0.777 | 13.333 | $1.65 \times 10^1$ | 0.00 |
| Group 2 – Group 1 | 7.000 | 14.925 | 0.777 | 13.333 | $1.65 \times 10^1$ | 0.00 |
| Group 2 – Group 1 | 8.000 | 14.925 | 0.777 | 13.333 | $1.65 \times 10^1$ | 0.00 |
| Group 2 – Group 1 | 9.000 | 14.925 | 0.777 | 13.333 | $1.65 \times 10^1$ | 0.00 |
| Group 2 – Group 1 | 10.000 | 14.925 | 0.777 | 13.333 | $1.65 \times 10^1$ | 0.00 |
| Group 2 – Group 1 | 11.000 | 14.925 | 0.777 | 13.333 | $1.65 \times 10^1$ | 0.00 |
| Group 2 – Group 1 | 12.000 | 14.925 | 0.777 | 13.333 | $1.65 \times 10^1$ | 0.00 |
| Group 3 – Group 1 | 1.000 | 18.992 | 0.815 | 17.323 | $2.07 \times 10^1$ | 0.00 |
| Group 3 – Group 1 | 2.000 | 20.730 | 0.802 | 19.086 | $2.24 \times 10^1$ | 0.00 |
| Group 3 – Group 1 | 3.000 | 22.468 | 0.793 | 20.844 | $2.41 \times 10^1$ | 0.00 |
| Group 3 – Group 1 | 4.000 | 24.206 | 0.785 | 22.598 | $2.58 \times 10^1$ | 0.00 |
| Group 3 – Group 1 | 5.000 | 25.944 | 0.780 | 24.346 | $2.75 \times 10^1$ | 0.00 |
| Group 3 – Group 1 | 6.000 | 27.682 | 0.778 | 26.089 | $2.93 \times 10^1$ | 0.00 |
| Group 3 – Group 1 | 7.000 | 29.420 | 0.778 | 27.827 | $3.10 \times 10^1$ | 0.00 |
| Group 3 – Group 1 | 8.000 | 31.158 | 0.780 | 29.560 | $3.28 \times 10^1$ | 0.00 |
| Group 3 – Group 1 | 9.000 | 32.896 | 0.785 | 31.288 | $3.45 \times 10^1$ | 0.00 |
| Group 3 – Group 1 | 10.000 | 34.634 | 0.793 | 33.010 | $3.63 \times 10^1$ | 0.00 |
| Group 3 – Group 1 | 11.000 | 36.372 | 0.802 | 34.728 | $3.80 \times 10^1$ | 0.00 |
| Group 3 – Group 1 | 12.000 | 38.110 | 0.815 | 36.441 | $3.98 \times 10^1$ | 0.00 |
| Group 3 – Group 2 | 1.000 | 4.067 | 0.832 | 2.362 | 5.77 | $3.77 \times 10^{-5}$ |
| Group 3 – Group 2 | 2.000 | 5.805 | 0.820 | 4.125 | 7.48 | $1.06 \times 10^{-7}$ |
| Group 3 – Group 2 | 3.000 | 7.543 | 0.810 | 5.882 | 9.20 | $4.59 \times 10^{-10}$ |
| Group 3 – Group 2 | 4.000 | 9.281 | 0.803 | 7.636 | $1.09 \times 10^1$ | $3.61 \times 10^{-12}$ |
| Group 3 – Group 2 | 5.000 | 11.019 | 0.798 | 9.384 | $1.27 \times 10^1$ | $5.11 \times 10^{-14}$ |
| Group 3 – Group 2 | 6.000 | 12.757 | 0.796 | 11.127 | $1.44 \times 10^1$ | $1.33 \times 10^{-15}$ |
| Group 3 – Group 2 | 7.000 | 14.495 | 0.796 | 12.865 | $1.61 \times 10^1$ | 0.00 |
| Group 3 – Group 2 | 8.000 | 16.233 | 0.798 | 14.598 | $1.79 \times 10^1$ | 0.00 |
| Group 3 – Group 2 | 9.000 | 17.971 | 0.803 | 16.326 | $1.96 \times 10^1$ | 0.00 |
| Group 3 – Group 2 | 10.000 | 19.709 | 0.810 | 18.049 | $2.14 \times 10^1$ | 0.00 |
| Group 3 – Group 2 | 11.000 | 21.447 | 0.820 | 19.767 | $2.31 \times 10^1$ | 0.00 |
| Group 3 – Group 2 | 12.000 | 23.185 | 0.832 | 21.481 | $2.49 \times 10^1$ | 0.00 |

Appendix Table A2. Comparison of candidate linear mixed-effects models (AIC, BIC, log-likelihood).

| Table A3. 12-week body-weight gains | | | | |
|---|---|---|---|---|
| Group | Estimate | Std_Error | Lower_95CI | Upper_95CI |
| Group 1: wild-type | 3.702 | 0.277 | 3.158 | 4.25 |
| Group 2: ob/ob pair-fed | 3.702 | 0.277 | 3.158 | 4.25 |
| Group 3: ob/ob unrestricted | 22.820 | 0.402 | 22.032 | $2.36 \times 10^1$ |

Appendix Table A3. Fixed-effect estimates for Models 1–3 ($\beta$, standard error, 95\% CI, p-value).

# 1.3 A3. Derivation of contrasts for group differences and weight gain

To make the link between scientific questions and model parameters fully explicit, this section shows the algebraic derivation of all linear contrasts used in the analysis. Starting from the mean structure of Model 3,

$$\mu_1(t)=\beta_0+\beta_1 t,\ \mu_2(t)=\beta_0+\beta_1 t+\beta_2,\ \mu_3(t)=\beta_0+\beta_1 t+\beta_3+\beta_4 t,$$

we derive the following weekly group differences:

$$\mu_2(t)-\mu_1(t)=\beta_2,\ \mu_3(t)-\mu_1(t)=\beta_3+\beta_4 t,\ \mu_3(t)-\mu_2(t)=(\beta_3-\beta_2)+\beta_4 t.$$

For 12-week mean weight gain, we define for each group $g$

$$\text{gain}_g=\mu_g(12)-\mu_g(1).$$

Under Model 3, this yields

$$\text{gain}_1 = 11\beta_1, \text{gain}_2 = 11\beta_1, \text{gain}_3 = 11(\beta_1+\beta_4), \text{gain}_3 - \text{gain}_1 = 11\beta_4.$$

For each estimand, Appendix Table A4 records the corresponding contrast used in `gmodels::estimable()` within `LMM_mouse_weight.R` in the parameter order

$$(\beta_0, \beta_1, \beta_2, \beta_3, \beta_4).$$

Appendix Table A4. Linear contrasts used for weekly group differences and 12-week mean weight gains.

## 1.4 A4. Complete weekly group differences

The main text reports summary values for the group differences at representative weeks. Appendix Table A5 provides the full set of estimates, standard errors, 95\% confidence intervals, and p-values for all weeks 1–12, for each of the three pairwise comparisons:

- Group 2 vs group 1 (pair-fed *ob/ob* vs wild-type)
- Group 3 vs group 1 (unrestricted *ob/ob* vs wild-type)
- Group 3 vs group 2 (unrestricted vs pair-fed *ob/ob*)

These values are exported from the `diff_tab_all` object created in `LMM_mouse_weight.R` and correspond to Figure 3 in the main text.

Appendix Table A5. Model-based differences in mean body weight between groups at each week (estimates, standard errors, 95\% CI, p-values).

## 1.5 A5. Additional diagnostic plots

The main manuscript presents the key diagnostic plots for the final model (Model 3), including:

- Figure 5 (main text): Residuals vs fitted values and residuals vs time.
- Figure 6 (main text): Normal Q–Q plot of marginal residuals and empirical distribution of random intercepts.

Appendix Figures A1–A4 provide further diagnostic detail:

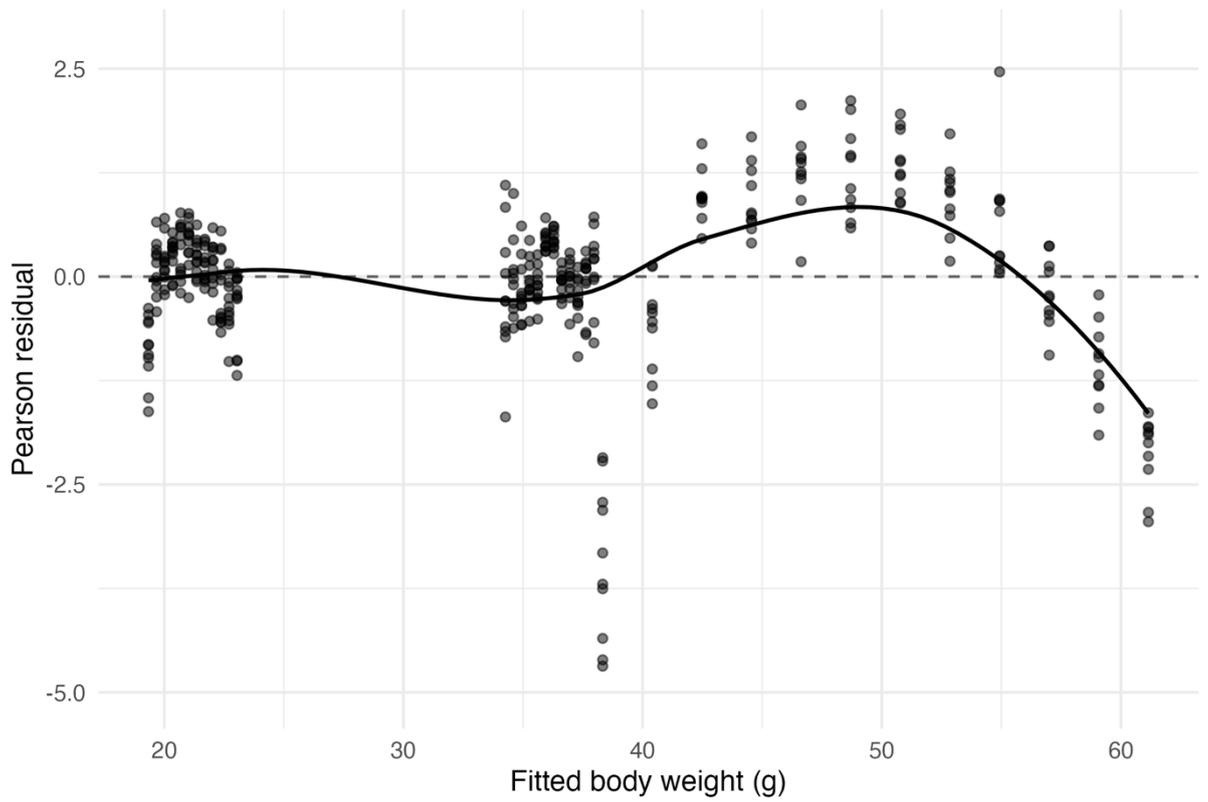

Figure 5. Residuals vs fitted values stratified by group (panelled by `grp`).

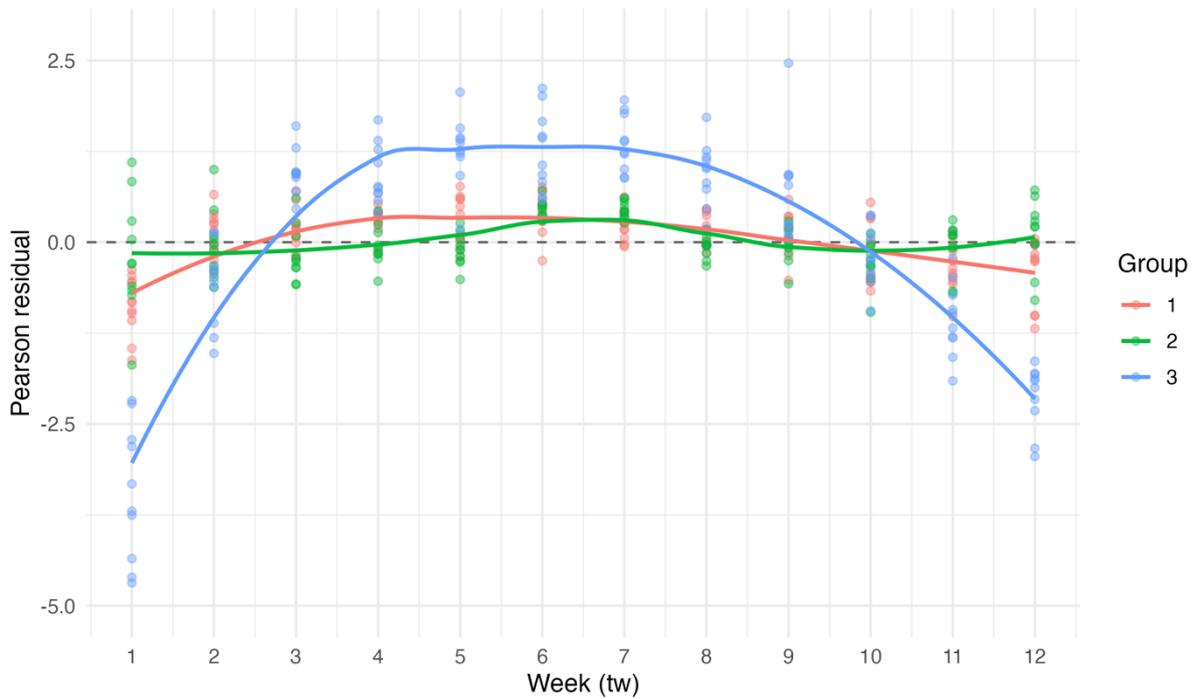

Figure 7. Residuals vs time stratified by group (panelled by `grp`).

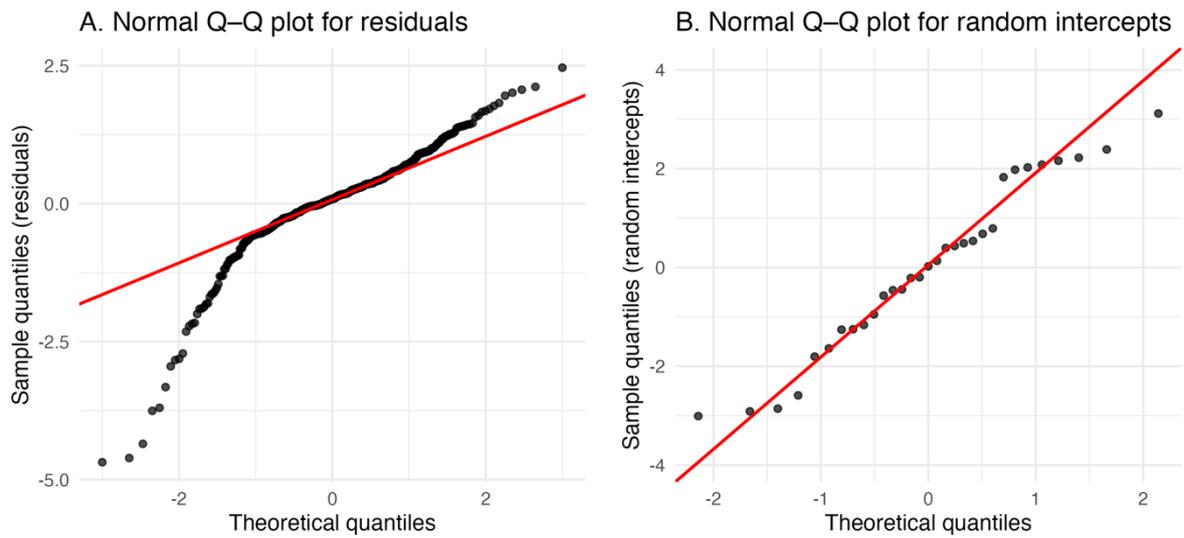

Figure 6. Normal Q–Q plot of random intercepts, with point labels for mouse IDs to help identify outliers or influential mice.

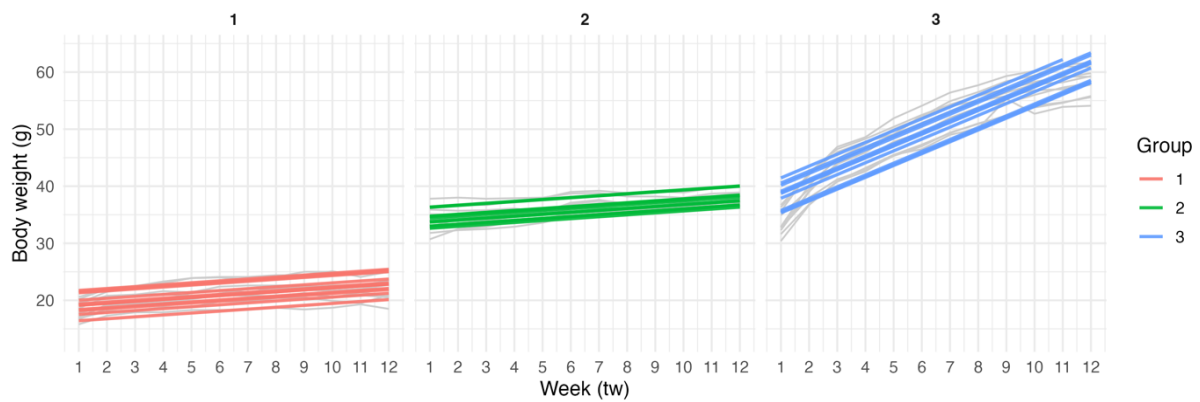

Figure S1. Fitted trajectories overlaid on individual mouse profiles for each group, illustrating how well Model 3 captures within-mouse trends.

All diagnostic figures were generated by the script `LMM_mouse_weight_diagnostics.R` and use consistent axis labels ("Fitted weight (g)", "Marginal residual (g)", "Week (tw)") and full figure captions so that they can be interpreted independently of the main text.

## 1.6 A6. Full sensitivity-analysis results

Section 7.6 of the main text summarises the sensitivity analyses comparing the baseline random-intercept model with alternative random-effects and correlation structures (random slope for time; autoregressive AR(1) within-mouse correlation; combined random intercept and slope). Appendix Table A6 presents detailed model fit indices (AIC, BIC, log-likelihood) and variance–covariance estimates for all sensitivity models fit
in `LMM_mouse_weight_sensitivity.R`.

Appendix Table A7 reports the re-estimated 12-week mean weight gains and key pairwise differences under each alternative model, allowing readers to verify the robustness of the main inferential conclusions (for example, the approximately 19 g excess gain for unrestricted *ob/ob* mice vs the other groups). For completeness, Appendix Figures A5–A6 show residual and random-effects diagnostics for the most complex model (random intercept + random slope), confirming that the additional complexity does not materially improve fit or change the substantive interpretation.

| TableS1_sensitivity_model_comparison_AIC_BIC_logLik | | | |
|---|---:|---:|---:|
| Model comparison: main vs sensitivity structures | | | |
| Model | AIC | BIC | logLik |
| Main: random intercept | 1,397.676 | 1,425.108 | −691.838 |
| RS: random intercept + slope | 1,401.676 | 1,436.946 | −691.838 |
| AR(1): rand. intercept + AR(1) | 1,135.029 | 1,166.380 | −559.514 |
| HV: rand. intercept + hetero. variance | 1,151.415 | 1,186.685 | −566.708 |

Appendix Table FigS1. AIC, BIC, log-likelihood, and variance–covariance components for baseline and sensitivity models.(check the github link)

| TableS2_fixed_effects_across_models | | | |
|---|---|---:|---:|
| Fixed-effect estimates and standard errors across models | | | |
| Model | Term | Estimate | Std_Error |
| Main | (Intercept) | 19.004 | 0.561 |
| Main | tw | 0.337 | 0.025 |
| Main | grp2 | 14.925 | 0.777 |
| Main | grp3 | 17.254 | 0.829 |
| Main | tw:grp3 | 1.738 | 0.044 |
| RS | (Intercept) | 19.004 | 0.561 |
| RS | tw | 0.337 | 0.025 |
| RS | grp2 | 14.925 | 0.777 |
| RS | grp3 | 17.254 | 0.829 |
| RS | tw:grp3 | 1.738 | 0.044 |
| AR(1) | (Intercept) | 18.168 | 0.686 |
| AR(1) | tw | 0.384 | 0.055 |
| AR(1) | grp2 | 15.414 | 0.849 |
| AR(1) | grp3 | 14.759 | 1.057 |
| AR(1) | tw:grp3 | 1.841 | 0.097 |
| HV | (Intercept) | 18.995 | 0.522 |
| HV | tw | 0.338 | 0.012 |
| HV | grp2 | 14.925 | 0.748 |
| HV | grp3 | 17.262 | 0.868 |
| HV | tw:grp3 | 1.737 | 0.061 |

Appendix Table S2. Model-based 12-week mean weight gains and pairwise differences under alternative random-effects and correlation structures(check the github link)

### TableS2_fixed_effects_across_models
Fixed-effect estimates and standard errors across models

| Model | Term | Estimate | Std_Error |
|---|---|---|---|
| Main | (Intercept) | 19.004 | 0.561 |
| Main | tw | 0.337 | 0.025 |
| Main | grp2 | 14.925 | 0.777 |
| Main | grp3 | 17.254 | 0.829 |
| Main | tw:grp3 | 1.738 | 0.044 |
| RS | (Intercept) | 19.004 | 0.561 |
| RS | tw | 0.337 | 0.025 |
| RS | grp2 | 14.925 | 0.777 |
| RS | grp3 | 17.254 | 0.829 |
| RS | tw:grp3 | 1.738 | 0.044 |
| AR(1) | (Intercept) | 18.168 | 0.686 |
| AR(1) | tw | 0.384 | 0.055 |
| AR(1) | grp2 | 15.414 | 0.849 |
| AR(1) | grp3 | 14.759 | 1.057 |
| AR(1) | tw:grp3 | 1.841 | 0.097 |
| HV | (Intercept) | 18.995 | 0.522 |
| HV | tw | 0.338 | 0.012 |
| HV | grp2 | 14.925 | 0.748 |
| HV | grp3 | 17.262 | 0.868 |
| HV | tw:grp3 | 1.737 | 0.061 |

Appendix Figures S2. Diagnostics for the random intercept + random slope model(check the github link)